\documentclass[11pt]{article}
\usepackage{calc}
\usepackage{color}
\usepackage{graphicx}
\usepackage{float}
\usepackage{placeins}
\usepackage{url}

\usepackage{booktabs}
\usepackage{tabularx}
\usepackage{multirow}
\usepackage{makecell}
\usepackage{threeparttable}
\usepackage{wrapfig}
\usepackage[export]{adjustbox}

\usepackage{mathtools}
\usepackage{bm}
\usepackage[bbgreekl]{mathbbol}
\usepackage{siunitx}

\usepackage{algorithmic}
\usepackage{algorithm}

\usepackage[biblabel]{cite}
\usepackage{titleref}

\usepackage[left=0.5in,right=0.5in,top=0.5in,bottom=0.5in,includefoot,heightrounded]{geometry}
\usepackage{lscape}
\usepackage[margin=15pt,font=small,labelfont={bf,sf},justification=justified]{caption}
\usepackage{subcaption} 
\usepackage{titling}
\usepackage{authblk}
\usepackage[pagewise]{lineno}
\usepackage{soul}
\usepackage{blindtext}
\usepackage{lipsum}
\usepackage{easylist}
\usepackage{tikz}

\usepackage{listings}
\usepackage{xcolor}

\usepackage{hyperref}
\usepackage{cleveref}

\newif\ifbembo 
\newif\ifcharter
\newif\iferewhon
\newif\iflibertine
\newif\iflibertinealt
\newif\ifpalantino
\newif\iftimesnewroman

\bembofalse
\chartertrue
\erewhonfalse
\libertinefalse
\libertinealtfalse
\palantinofalse
\timesnewromanfalse

\ifbembo
  \usepackage[p,osf]{ETbb} 
  \usepackage[scaled=.95,type1]{cabin} 
  \usepackage[varqu,varl]{zi4}
  \usepackage[T1]{fontenc}
  \usepackage[libertine,vvarbb]{newtxmath}
  \usepackage[cal=boondoxo,bb=boondox]{mathalfa}
\fi

\ifcharter
  \usepackage[scaled=.98,p]{XCharter}
  \usepackage[scaled=1.04,varqu,varl]{inconsolata}
  \usepackage[type1]{cabin}
  \usepackage[uprightscript,xcharter,vvarbb,scaled=1.05]{newtxmath}
  \usepackage[cal=euler,scr=boondoxo,bb=boondox]{mathalpha}  
\fi

\iferewhon
  \usepackage[p,osf,scaled=.98,space]{erewhon} 
  \usepackage[varqu,varl]{inconsolata} 
  \usepackage[type1,scaled=.95]{cabin} 
  \usepackage[utopia,vvarbb]{newtxmath}
\fi

\iflibertine
  \usepackage[sb]{libertinus}
  \usepackage[T1]{fontenc}
  \usepackage{textcomp}
  \usepackage[varqu,varl]{zi4}
  \usepackage{libertinust1math} 
  \usepackage[scr=boondoxo,bb=boondox]{mathalpha} 
\fi

\iflibertinealt
  \usepackage[sb]{libertinus}
  \usepackage[T1]{fontenc}
  \usepackage{textcomp}
  \usepackage{libertinust1math} 
  \usepackage[scr=boondoxo,bb=boondox]{mathalpha} 
\fi

\ifpalantino
  \usepackage[largesc]{newpxtext}
  \usepackage[vvarbb]{newpxmath}
\fi

\iftimesnewroman
  \usepackage{newtxtext} 
  \usepackage[scaled=0.95]{inconsolata} 
  \usepackage[vvarbb]{newtxmath} 
\fi

\usepackage{microtype}
\pretitle{\begin{center}\bfseries\Large}
\posttitle{\end{center}}

\predate{\begin{center}\normalsize}
\postdate{\end{center}}

\pretitle{\begin{center}\bfseries\sffamily\LARGE}
\posttitle{\end{center}}
\usepackage{abstract}

\allowdisplaybreaks
\usepackage{sectsty} 
\allsectionsfont{\sffamily}

\makeatletter
\patchcmd{\LS@rot}{90}{-90}{}{}
\patchcmd{\endlandscape}{90}{-90}{}{}
\makeatother

\def \L2{L^2}

\definecolor{vargreen}{rgb}{0.0, 0.5, 0.0}
\definecolor{varpurp}{rgb}{0.5, 0.0, 0.5}

\renewcommand{\eqref}[1]{Eq.\,(\ref{#1})}
\newcommand{\eqrefs}[1]{Eqs.\,(\ref{#1})}

\title{
  AFSI: Automated Fluid-Structure Interaction Solver Development for Nonlinear Solid Mechanics
}
\author[1-3]{Pengfei ~Ma} 
\author[1-3$^{*}$]{Li ~Cai}
\author[1-3]{Xuan ~Wang}
\author[4]{Hao ~Gao}

\affil[1]{Xi'an Key Laboratory of Scientific Computation and Applied Statistics, China}
\affil[2]{NPU-UoG International Cooperative Lab for Computation and Application in Cardiology, China}
\affil[3]{School of Mathematics and Statistics, Northwestern Polytechnical University, China}
\affil[4]{School of Mathematics and Statistics, University of Glasgow, UK}
\affil[*]{\texttt{Corresponding author, caili@nwpu.edu.cn}}

\date{Manuscript generated on \today} 
\begin{document}
\maketitle

\begin{abstract}
    
    \textcolor{black}{AFSI} is a novel, open-source fluid–structure interaction (FSI) solver 
    that extends the capabilities of the FEniCS finite element library through 
    an immersed boundary (IB) framework. Designed to simulate large deformations 
    in hyperelastic materials—such as cardiac tissue—\textcolor{black}{AFSI} avoids the need for expensive remeshing by coupling a Lagrangian representation of the solid with an Eulerian description of the surrounding fluid. This approach retains the full expressiveness of FEniCS's variational formulations, function spaces, and time integration schemes.
    Implemented in a hybrid Python/C++ architecture, \textcolor{black}{AFSI} allows users to define geometries, constitutive models (e.g., the Holzapfel–Ogden law for myocardium), and strain energy functions directly in Python, while delegating performance-critical tasks such as assembly and linear solvers to optimized C++ backends. Its concise and modular Python API facilitates the setup of FSI simulations, enabling users to easily modify discretization strategies or analyze results using standard FEniCS post-processing tools.
    By combining the flexibility of FEniCS with a robust immersed boundary formulation, \textcolor{black}{AFSI} empowers rapid prototyping of complex nonlinear solid–fluid interaction problems, making it a powerful tool for simulating biomechanical systems and other applications involving highly deformable structures in flow.

\end{abstract}   

\newpage

\section{Introduction}
\label{sec::introduction}

Simulating the interactions between flexible solid materials and surrounding fluids is of 
fundamental importance across a wide range of engineering and biomedical applications. For 
instance, \textcolor{black}{evaluating the blood bumping function of ventricle}, 
the open and close of heart valves during cardiac 
cycles \cite{lee_fluidstructure_2020, bavo_fluid-structure_2016, lin_fluidstructure_2019, kuchumov_fluidstructure_2023}, 
and the motion of soft robotic swimmers \cite{borazjani_review_2015} or biomimetic underwater 
robots \cite{wang_development_2022, li_underwater_2023, qu_recent_2024} all require a 
detailed understanding of fluid-structure interaction (FSI). These systems typically 
involve large deformations, nonlinear material responses, and intricate fluid dynamics, 
posing significant challenges for both experimental measurement and numerical modeling 
\cite{pramanik_computational_2024, abbas_state_art_2022}. 

Physical experiments in FSI domains are often costly and constrained by limited spatial and temporal resolution.
As a result, accurate and efficient numerical FSI solvers have become indispensable tools for understanding and predicting the behavior of these complex coupled systems.
However, despite the availability of numerous scientific computing frameworks, practical FSI simulation still faces significant challenges.
The classical Arbitrary Lagrangian–Eulerian (ALE) method is known for its numerical stability and accuracy \cite{bergersen_turtlefsi_2020},
but it often requires generating conforming fluid meshes that move and 
deform with the solid geometry, which is an operation that becomes prohibitively expensive 
in three-dimensional simulations. 
As observed in the simulation of prosthetic aortic valves, the remeshing 
overhead in ALE-based FSI can far exceed the cost of solving the coupled system itself,
severely limiting usability in dynamic scenarios \cite{bavo_fluid_structure_2016}.
In contrast, the immersed boundary (IB) method employs nonconforming meshes between fluid and structure.
Although its application to materials with finite volume 
and nonlinear material responses developed later than ALE methods \cite{boffi_hyper_elastic_2008}, 
the IB method has rapidly advanced, particularly in computational biology, and is now widely used to simulate biological 
fluid dynamics, especially in cardiac mechanics \cite{feng_whole_heart_2024}.
It supports nonconforming discretizations of fluid and structure, eliminating the need for dynamically generated body-fitted meshes.
This is especially advantageous in problems involving large deformations, structural displacements, or contact interactions \cite{griffith_hybrid_2017}.
Although many IB-based FSI solvers exist, only a few support a wide range of hyperelastic material models.
IBAMR is one of the most sophisticated IB-based solvers. However, 
its complexity and heavy reliance on external libraries 
such as SAMRAI, PETSc, and libMesh—combined with 
its extensive low-level C++ codebase—pose a steep 
learning curve for new users.


In our work, we address these challenges
through a combination of tailored numerical methods and efficient software design. 
We adopt a nodal immersed boundary finite element (IBFE) method \cite{wells_nodal_2023}, 
which decomposes the FSI problem into an Eulerian Navier-Stokes solver for the 
fluid and a Lagrangian solid solver. 
The coupling between these two domains is achieved via 
integral transforms that transfer information between Eulerian and Lagrangian variables.
A key contribution of our framework is the C++ implementation of these coupling operators, 
which uses shared-memory parallelism via Intel TBB, and is exposed to Python through 
nanobind. This design allows users to write simulations in compact, 
human-readable Python code, typically within a few hundred lines 
per example, where each line closely mirrors the corresponding 
mathematical expression. Users do not need to manage or 
even be aware of the underlying parallelism, which is fully abstracted away.
Besides, it is integrated into FEniCS, which translates high-level variational 
forms into optimized C++ code at runtime. This eliminates the need for manual 
low-level finite element coding and significantly streamlines the 
development process. FEniCS also provides built-in support for parallel computing, 
enabling efficient large-scale simulations.
The framework is designed to be modular and extensible, 
allowing flexible experimentation with different finite element discretizations, 
time-stepping schemes, and nonlinear material models. 
Its effectiveness is demonstrated through several 
numerical experiments, including blood flow-driven motion of 
a two-dimensional valve and torsion of an elastic beam immersed in fluid. 
These examples showcase the solver's ability to handle complex geometries, 
large deformations, and nonlinear constitutive behaviors.

The remainder of this paper is structured as follows.
Section 2 presents the governing equations of the coupled fluid-structure system, 
hybrided with Eulerian and Lagrangian frameworks, constitutive models for 
hyperelastic materials, and the nodal IBFE method formulation.
Section 3 describes the implementation of the solver, with a focus on temporal 
discretization, the automated finite element framework, and its hybrid Python/C++ architecture.
Section 4 demonstrates the solver's accuracy and robustness through benchmark cases.
Finally, Section 5 summarizes the main contribution of this study 
and discusses future research directions. 

\section{Methodology}\label{sec2}

\subsection{Fluid-structure interaction problem}
This study is based on the IB method derived by Boffi et al.\cite{boffi2008hyper}, which is used to solve FSI systems. The method is mathematically described by the following system:
\begin{subequations}\label{IB::continuous}
    \begin{align}
        \rho \left(\frac{\partial \mathbf{u}}{\partial t}(\mathbf{x}, t) + \mathbf{u}(\mathbf{x},t) \cdot \nabla \mathbf{u}(\mathbf{x},t)\right)
        &+ \nabla p(\mathbf{x}, t) - \mu \Delta \mathbf{u}(\mathbf{x}, t)
        = \mathbf{f}(\mathbf{x}, t), && \text{in } \Omega \times [0, T], \label{IB::momentum} \\
        \nabla \cdot \mathbf{u}(\mathbf{x}, t)
        &= 0, && \text{in } \Omega \times [0, T], \label{IB::incompressible} \\
        \int_{B_r} \mathbf{F}(\mathbf{X}, t) \cdot \mathbf{V}(\mathbf{X}) \, d\mathbf{X}
        &= -\int_{B_r} \mathbb{P}(\mathbf{X}, t) : \nabla_{\mathbf{X}} \mathbf{V}(\mathbf{X}) \, d\mathbf{X}, && \text{in } B_r \times [0, T], \label{IB::solid::weak} \\
        \mathbf{f}(\mathbf{x}, t)
        &= \int_{B_r} \mathbf{F}(\mathbf{X},t) \, \delta\left(\mathbf{x}-\mathcal{X}(\mathbf{X},t)\right) \, d\mathbf{X}, && \text{in } \Omega \times [0, T], \label{IB::force::spread} \\
        \frac{\partial \mathcal{X}(\mathbf{X}, t)}{\partial t}
        &= \mathbf{U}(\mathbf{X},t) = \int_{\Omega} \mathbf{u}(\mathbf{x},t) \, \delta\left(\mathbf{x}-\mathcal{X}(\mathbf{X},t)\right) \, d\mathbf{x}, && \text{in } B_r \times [0, T]. \label{IB::velocity::interp}
    \end{align}
    \end{subequations}
    The variables and components in the above system are defined as follows:
\begin{itemize}
    \item  $\rho$ is the constant density shared by the fluid and structure.
    \item  The velocity $\mathbf{u}(\mathbf{x}, t)$ and pressure $p(\mathbf{x}, t)$ are defined on the fixed Eulerian computational domain $\Omega$ and are governed by \eqref{IB::momentum}–\eqref{IB::incompressible}.  
    \item  $\mu$ is the dynamic viscosity of the incompressible Newtonian fluid.
    \item The external force density $\mathbf{f}(\mathbf{x}, t)$ acts as a coupling term between the Lagrangian and Eulerian frameworks (\eqref{IB::force::spread}).
    \item The Lagrangian elastic force density $\mathbf{F}(\mathbf{X}, t)$, defined on the reference configuration $B_r$ of the immersed structure, satisfies the weak formulation in \eqref{IB::solid::weak} for all smooth test functions $\mathbf{V}(\mathbf{X})$. Here, $\mathbb{P}(\mathbf{X}, t)$ is the first Piola–Kirchhoff stress tensor.
    \item The structure's motion is described by its current position $\mathcal{X}(\mathbf{X}, t)$ and velocity $\mathbf{U}(\mathbf{X}, t)$, which are related to the surrounding fluid velocity via interpolation using the Dirac delta function (\eqref{IB::velocity::interp}).
    \item The Eulerian spatial coordinates are denoted by $\mathbf{x} = (x_1, \dots, x_d) \in \Omega$, and the Lagrangian coordinates by $\mathbf{X} = (X_1, \dots, X_d) \in B_r$.
    \item The Dirac delta function $\delta(\mathbf{x} - \mathcal{X}(\mathbf{X}, t))$ serves to mediate the two-way coupling between fluid and structure by enabling force spreading and velocity interpolation between the Eulerian and Lagrangian frames.
\end{itemize}

The computational model is subject to the following initial and boundary conditions:
\begin{equation}\label{eq::IB::continuous::boundary::intial}
    \begin{aligned}
        \mathbf{u}(\mathbf{x}, t)&=\mathbf{w}_D(\mathbf{x},t), && \text { on } \partial \Omega_D \times[0, T], \\ 
        \mu\frac{\partial\mathbf{u}(\mathbf{x},t)}{\partial\mathbf{n}}-p(\mathbf{x},t)\mathbf{n}&=\mathbf{w}_{N}(\mathbf{x},t),&&\text { on } \partial\Omega_{N} \times[0, T],\\
        \mathbf{u}(\mathbf{x}, 0)&=\mathbf{u}_0(\mathbf{x}), && \text { in } \Omega .
    \end{aligned}
\end{equation}
where $\partial\Omega=\partial\Omega_D\cup\partial\Omega_N$ is the boundary of the fluid domain, 
$\mathbf{w}_D(\mathbf{x}, t)$ is the prescribed velocity on the boundary, 
$\mathbf{w}_N(\mathbf{x}, t)$ is the prescribed traction on the boundary, 
and $\mathbf{u}_0(\mathbf{x})$ is the initial velocity field.

\subsection{Mechanical properties of the immersed structure}

This paper focuses on hyperelastic structures, whose material response is described by a strain energy density function 
$\Psi(\mathbb{F}(\mathbf{X},t))$. Accordingly, the first Piola–Kirchhoff stress tensor $\mathbb{P}(\mathbf{X}, t)$ is computed as
\begin{align*}
\mathbb{P}(\mathbf{X}, t) = \frac{\partial \Psi(\mathbb{F}(\mathbf{X}, t))}{\partial \mathbb{F}(\mathbf{X}, t)},
\end{align*}
where $\mathbb{F}(\mathbf{X},t)$ is the deformation gradient, defined by
\begin{align*}
\mathbb{F}(\mathbf{X}, t) = \frac{\partial \mathcal{X}(\mathbf{X}, t)}{\partial \mathbf{X}},
\end{align*}
and its determinant is denoted by $J(\mathbf{X}, t) = \det\left(\mathbb{F}(\mathbf{X}, t)\right)$.

Since the solid materials considered in this study are incompressible or nearly incompressible, 
we employ a dilatational penalty method to model their mechanical behavior (see \cite{vadala2020stabilization, gerhard2000nonlinear}). 
In this approach, the strain energy density function $\Psi(\mathbb{F})$ is decomposed into two parts: an isochoric component 
and a volumetric component:
$$
    \Psi(\mathbb{F}) = W_{\text{iso}}(\mathbb{F}) + U_{\text{vol}}(J),
$$
representing the volume-preserving and volume-changing elastic response of the material, respectively.


\section{Implementation}\label{sec3}
\subsection{Discretization}\label{sec3:1}

For the system \eqref{IB::continuous}, the time interval $[0, T]$ is uniformly divided into 
$N$ non-overlapping subintervals $(t_n, t_{n+1}]$, where $n = 1, 2, \dots, N$, and the time 
step size is defined as $\Delta t = t_n - t_{n-1}$. The notation $(\cdot)^n$ denotes the value of 
a variable at time step $t_n$. some varibales use the n+1, some variables use n.

\textcolor{black}{
The background domain is triangulated using quadrilateral elements for two-dimensional problems and hexahedral elements for three-dimensional problems. 
The velocity field is approximated using $Q2$ finite elements, while the pressure field is approximated using $Q1$ elements. 
The interpolation nodes for the finite element basis functions are illustrated in Figures 1 and 2.
}

\textcolor{black}{
The structural domain is triangulated using triangular elements for two-dimensional problems and tetrahedral elements for three-dimensional problems.
The structural force field $\mathbf{F}(\mathbf{X}, t)$ and current position field $\mathcal{X}(\mathbf{X}, t)$ are approximated using $P2$ finite elements.
The interpolation nodes for the finite element basis functions are illustrated in Figures 3 and 4.
}

\subsection{Automated finite element solution}
The FEniCS framework provides a convenient way to express strain energy 
density functions in symbolic form, maintaining a close correspondence with their mathematical notation.
For example, the strain energy density function of a neo-Hookean material is given by
$$
W=\frac{1}{2}\mu_s(I_1-d)-\mu_s\ln(J)+\frac{1}{2}\lambda\ln(J)^2,
$$
where the notations have been defined above. This expression can be implemented in Python within FEniCS as follows.
\begin{lstlisting}[language=Python]
def NeoHookeanModel(F, mu_s, lmbda, mesh):
    d = mesh.geometry().dim()  
    C = F.T*F                 
    J = det(F)                 
    I1 = tr(C)                 
    W = (mu_s/2)*(I1 - d) - mu_s*ln(J) + (lmbda/2)*(ln(J))**2
    return W
\end{lstlisting}
From this symbolic form, the first Piola-Kirchhoff stress tensor in the weak formulation \eqref{IB::solid::weak}
can be derived automatically in FEniCS simply by calling "diff(W, F)". As the discretization method defined in subsection \ref{sec3:1}, 
FEniCS generates and compiles highly optimized C++ code at runtime, ensuring both clarity in implementation and efficiency in computation.

As for the solver for the incompressible Navier-Stokes equations, \eqref{IB::momentum} and \eqref{IB::incompressible}, 
the discretization approach has already been described in Subsection~\ref{sec3:1}. As discussed in \cite{Langtangen2012}, various implementations are possible; 
in our work, unless stated otherwise, we adopt Chorin's projection method.

\subsection{Eulerian-Lagrangian coupling}
The Eulerian-Lagrangian coupling is implemented through the force spreading (\eqref{IB::force::spread}) 
and velocity interpolation (\eqref{IB::velocity::interp}) operators. While these formulations are not 
natively supported in FEniCS due to their requirement for interaction between distinct 
finite element spaces, we have developed a novel framework that integrates 
this capability directly within FEniCS, enabling complete solution of 
the immersed boundary system \eqrefs{IB::continuous}. This integration represents a key contribution of our work.

Our implementation builds upon the nodal coupling scheme introduced by \cite{wells_nodal_2023}, 
with adaptations for finite element discretization. 
In \cite{wells_nodal_2023}, the scheme is implemented using a finite difference method with 
Eulerian interaction nodes defined on a staggered grid. 
In the present work, the coordinates of the degrees of freedom illustrated in 
\textcolor{black}{Figures 1 and 2} chosen as the Eulerian interaction nodes. 
The Lagrangian interaction nodes follow the arrangement in \textcolor{black}{Figures 3 and 4}, as in \cite{wells_nodal_2023}.

The computational core is implemented in C++ for optimal performance, 
with Python interfaces exposed via nanobind \cite{nanobind}. 
This hybrid approach preserves computational efficiency while maintaining compatibility with FEniCS's Python ecosystem. 
It enables seamless integration with FEniCS's existing infrastructure and user-friendly Python interface. 
The implementation provides two key classes:
\begin{itemize}
\item \texttt{IBMesh}: Creates structured background meshes and manages DoF mappings for efficient coupling operations;
\item \texttt{IBInterpolation}: Handles the core coupling operations between Eulerian and Lagrangian domains.
\end{itemize}
They can be initialized using the following code.
\begin{lstlisting}[language=Python]
eulerian_mesh = IBMesh([Point(0, 0), Point(1, 1)], [10, 10], order_velocity)
lagrangian_mesh = RectangleMesh(Point(0.2, 0.2), Point(0.8, 0.8), 10, 10)
ib = IBInterpolation(eulerian_mesh, lagrangian_mesh)
\end{lstlisting}
The key coupling operations of \eqref{IB::force::spread} and \eqref{IB::velocity::interp} are implemented through the following interface.
\begin{lstlisting}[language=Python]
ib.fluid_to_solid(flow_velocity._cpp_object, solid_velocity._cpp_object)
ib.solid_to_fluid(flow_force._cpp_object, solid_force._cpp_object)
\end{lstlisting}

\subsection{Parallelism}

\cite{wang_massively_2020}
Gathered on the same rank

currently the bottleneck of the entire problem

\section{Examples}\label{sec4}
To demonstrate the applicability and reliability of the developed software AFSI in addressing FSI problems 
in biomechanics, two representative numerical benchmark cases are examined in this section.
The first is a two-dimensional heart valve model, designed to evaluate the software's capability in 
handling strongly coupled interactions between thin structures and surrounding fluid. 
The second is a three-dimensional idealized left ventricle model, 
aimed at assessing the stability and accuracy of AFSI under complex geometries and nonlinear coupling conditions. 
It is important to note that these benchmark cases are intended solely for research validation and will not 
be included in the public release of the software. Instead, a set of simplified and reproducible example 
cases will be provided at release, serving as references for users to understand the core functionalities 
and basic usage of AFSI.

\subsection{Idealized heart valve}

\begin{figure}[!ht]
    \centering
    \includegraphics[width=0.9\textwidth]{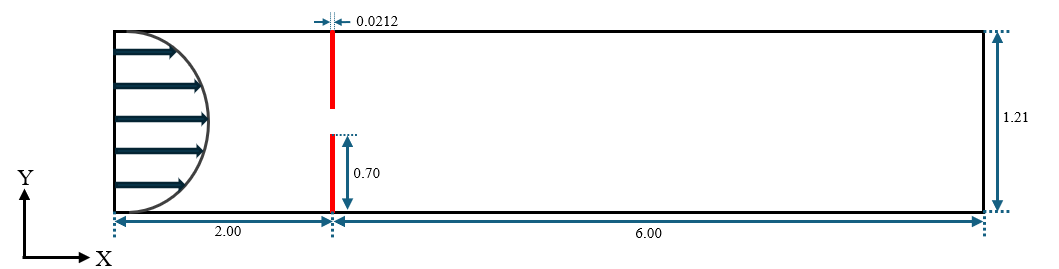}
    \caption{Setup of the idealized heart valve model. The black outline denotes the fixed 
    computational domain $\Omega$, while the red regions represent the deformable valve leaflets.}
    \label{fig::Idealized_heart_valve::2D}
\end{figure}

Here, a classical benchmark problem of an idealized two-dimensional heart valve is adopted, which has been extensively investigated and discussed 
in the literature \cite{black2025immersed, kamensky2015immersogeometric, hesch2012continuum, gil2010immersed}.
Based on the setup proposed in \cite{black2025immersed, kamensky2015immersogeometric}, a thin cantilever beam—representing a valve leaflet—was 
fixed to each of the upper and lower walls of a channel filled with an incompressible Newtonian 
fluid, as shown in Figure \ref{fig::Idealized_heart_valve::2D}. 
This configuration was used to assess the accuracy of AFSI in capturing heart valve dynamics 
under pulsatile flow conditions. To avoid structural contact, the beams were intentionally designed 
to be slightly shorter than the distance between the channel centerlines. 
This setup can also be interpreted as a simplified model of a check valve.

For this problem, the boundary conditions are set as follows: The inlet is located on the left boundary of the domain $\Omega$, 
    where the fluid is driven by a time-dependent velocity profile  
    \begin{align*}
    \mathbf{v}(\mathbf{x}, t)= 
    \begin{cases}
        5(\sin (2 \pi t)+1.1) y(1.61-y) \mathbf{e}_x, & t>0, \\ 
        \mathbf{0}, & \text { otherwise },\end{cases}
    \end{align*}
in order to replicate the characteristics of pulsatile flow. No-slip boundary conditions are imposed on the upper and lower walls of the domain, 
corresponding to zero velocity ($\mathbf{u} = \mathbf{0}$). 
The right boundary serves as the outlet and is assigned a free traction condition to allow the fluid to exit 
the domain freely. 
    
Under the same geometry and boundary conditions, 
we next analyze two valve materials with different constitutive models: isotropic and anisotropic.
\subsubsection{FSI validation: leaflets with isotropic constitutive Materials}
The first case considers the leaflets as isotropic hyperelastic materials, 
modeled using the Neo-Hookean model with a dilatational penalty term, and the model is defined as follows:

\begin{align*}
    \Psi=\frac{\mu^{s}}{2}\left(\text{tr}(\mathbb{C}-3)\right)-\mu^{s}\text{ln}(J)+\frac{\lambda^{s}}{2} \left(\text{ln}(J)\right)^2,
\end{align*}
where $\mu^{s}=\frac{E}{2(1+\nu)}$ and $\lambda^{s}=\frac{E\nu}{(1+\nu)(1-2\nu)}$ are the Lam\'{e} parameters. 
The parameter settings used for the simulation of this model are listed in Table \ref{tab::heart::valve::params::isotropic}.
It should be noted that the constitutive model originally represents a compressible material; however, 
due to the chosen Poisson’s ratio of $\nu = 0.4$, the material is considered nearly incompressible in practice.
\begin{table}[!h]
    \centering
    \begin{tabular}{l|c|c}
        \toprule
        \textbf{Symbol} & \textbf{Value} & \textbf{Unit}  \\
        \midrule
        $T$     & 3.0   & $\text{s}$ \\
        $\rho$  & 100   & $\frac{\text{g}}{\text{cm}^3}$ \\
        $\mu$   & 10.0  & $\text{g}/(\text{cm}\cdot\text{s})$\\
        $E$  & $5.6\times 10^{-7}$  & $\frac{\text{dyn}}{\text{cm}^2}$\\
        $\nu$  & 0.4 &  \\
        $\Delta t$ & $1\times 10^{-4}$ & $\text{s}$\\
        \bottomrule
    \end{tabular}
    \caption{Parameters for the idealized heart valve model with isotropic properties.}
    \label{tab::heart::valve::params::isotropic}
\end{table}

\begin{figure}[!h]
    \raggedright
    \begin{subfigure}[b]{0.48\textwidth}
        \centering
        \includegraphics[height=0.29\textwidth]{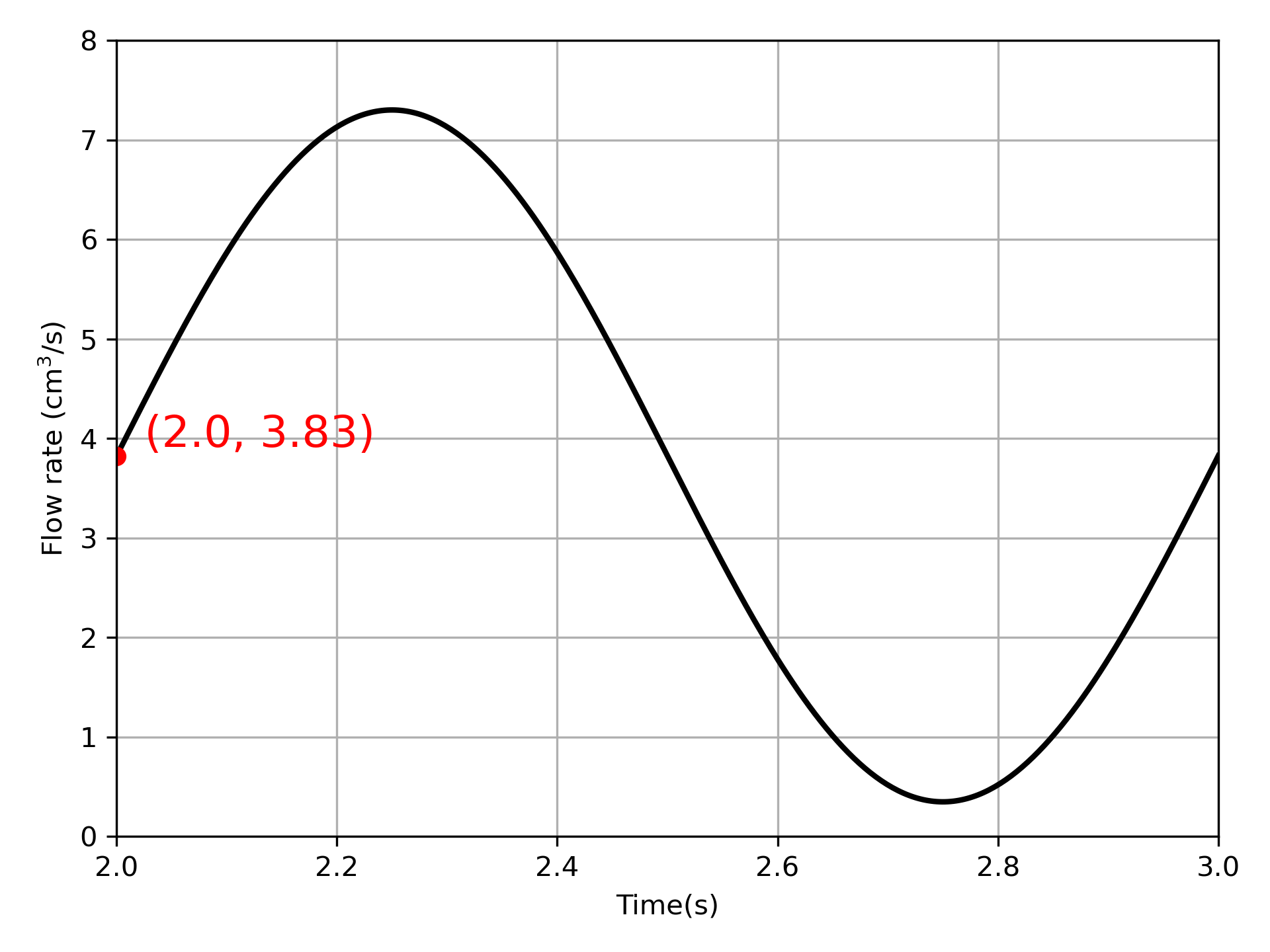}
        \caption{2.0~\text{s}}
        \label{fig::aortic_valve::streamlines::velocity::1}
    \end{subfigure}
    \hspace{-2.7em}
    \begin{subfigure}[b]{0.48\textwidth}
        \centering
        \includegraphics[height=0.29\textwidth]{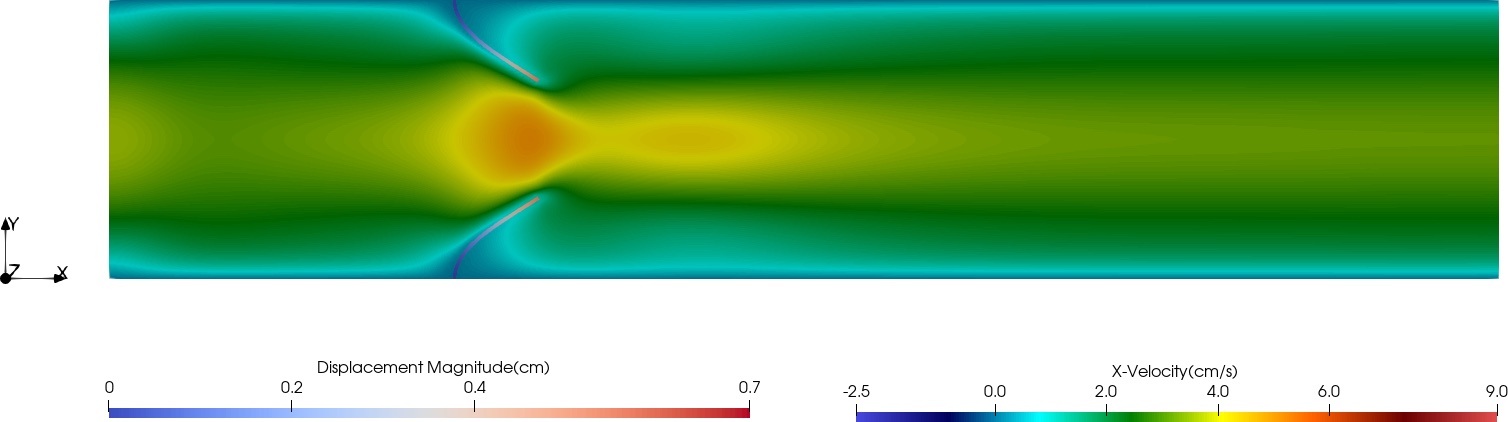}
        \caption{2.0~\text{s}}
        \label{fig::aortic_valve::streamlines::pressure::1}
    \end{subfigure}
    \hspace{-2.7em}
    \begin{subfigure}[b]{0.48\textwidth}
        \centering
        \includegraphics[height=0.29\textwidth]{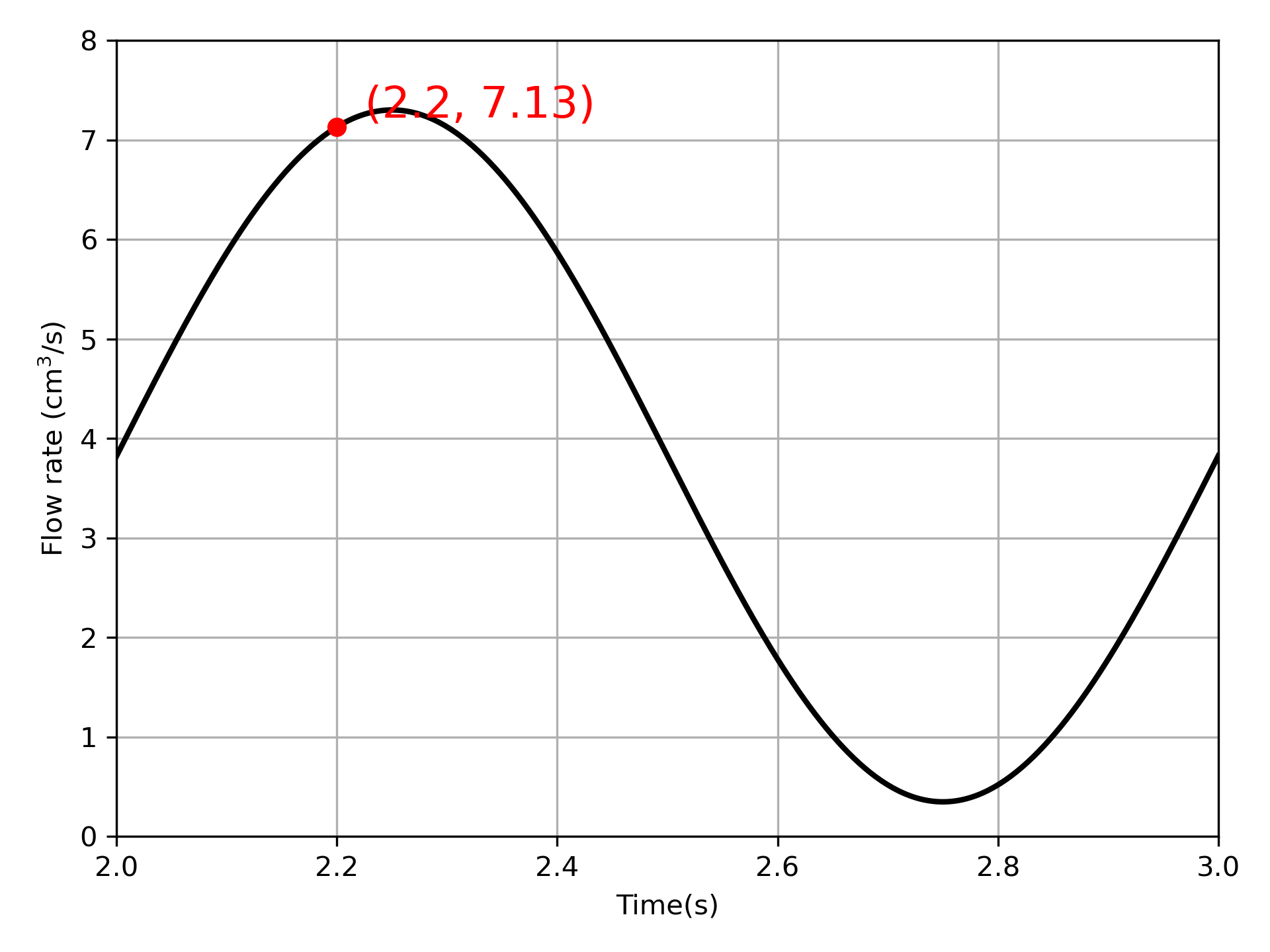}
        \caption{2.2~\text{s}}
        \label{fig::aortic_valve::streamlines::velocity::2}
    \end{subfigure}
    \hspace{-2.7em}
    \begin{subfigure}[b]{0.48\textwidth}
        \centering
        \includegraphics[height=0.29\textwidth]{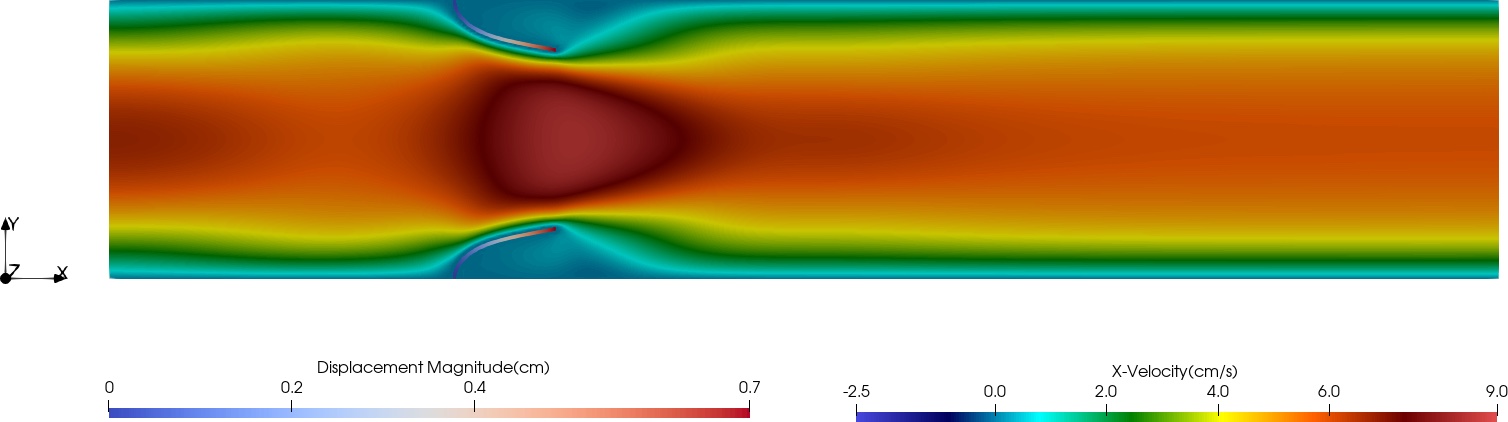}
        \caption{2.2~\text{s}}
        \label{fig::aortic_valve::streamlines::pressure::2}
    \end{subfigure}
    \hspace{-2.7em}
    \begin{subfigure}[b]{0.48\textwidth}
        \centering
        \includegraphics[height=0.29\textwidth]{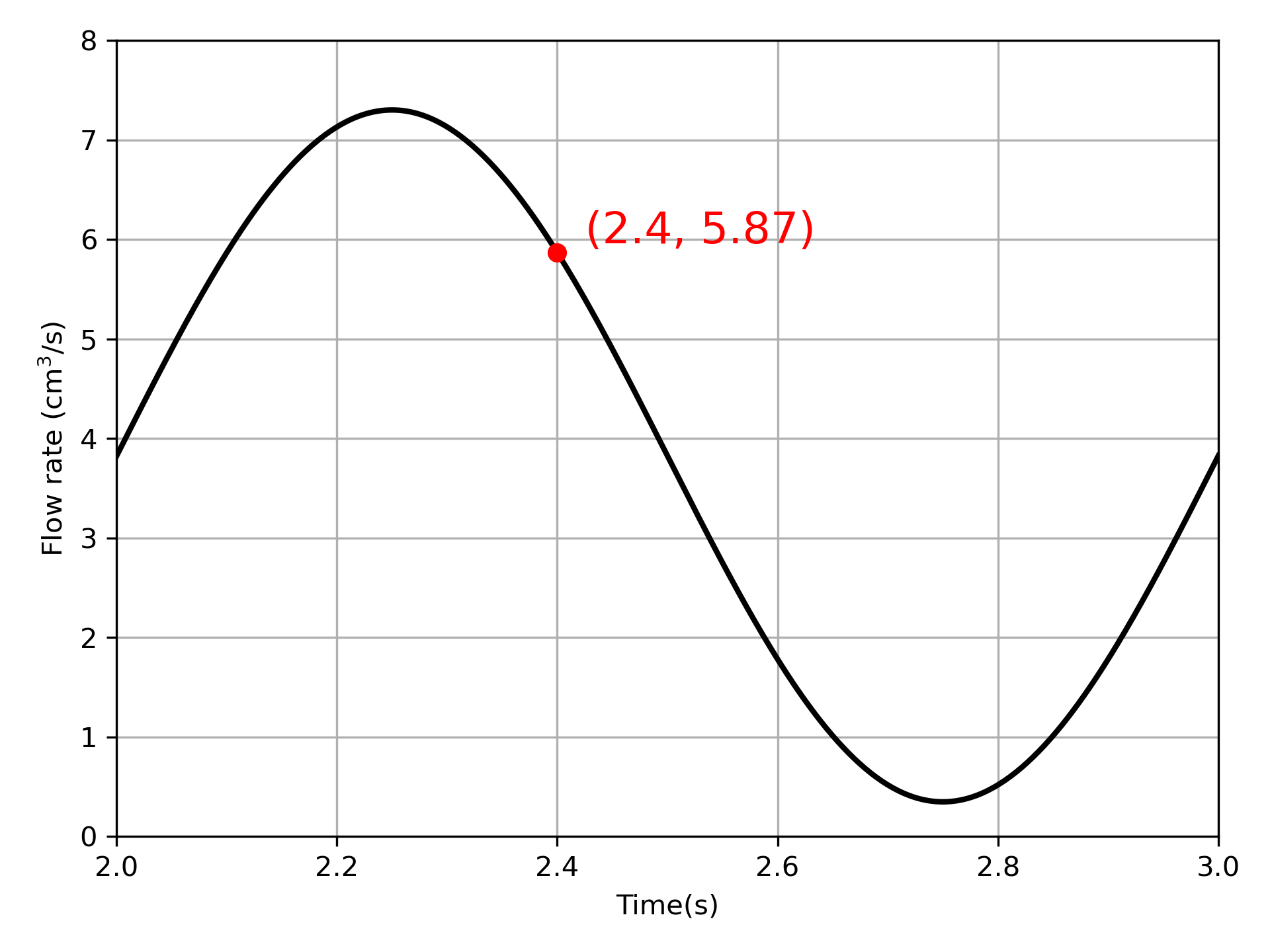}
        \caption{2.4~\text{s}}
        \label{fig::aortic_valve::streamlines::velocity::3}
    \end{subfigure}
   \hspace{-2.7em}
    \begin{subfigure}[b]{0.48\textwidth}
        \centering
        \includegraphics[height=0.29\textwidth]{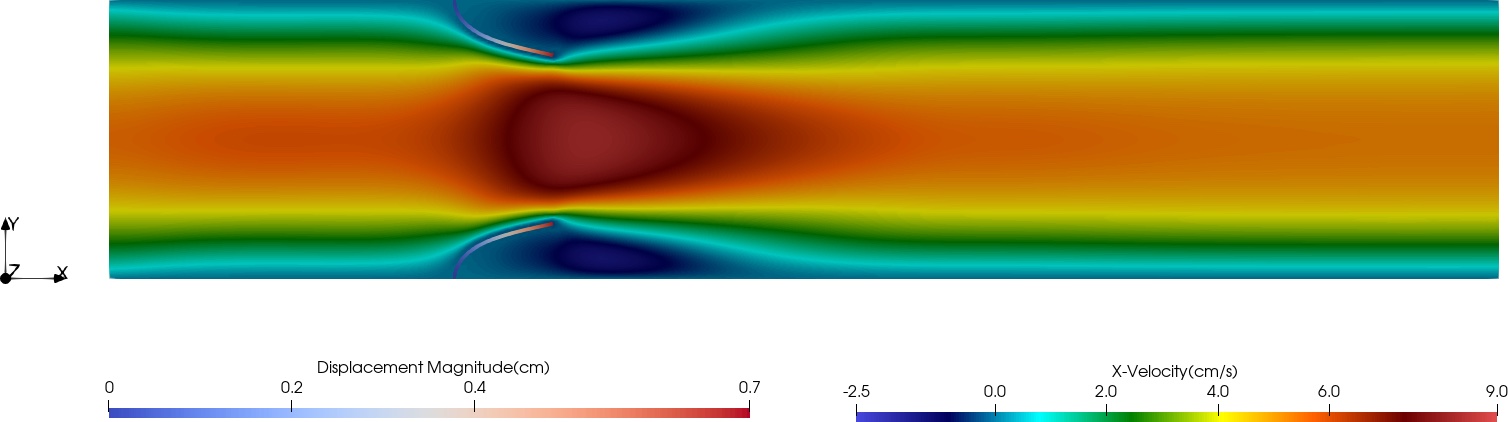}
        \caption{2.4~\text{s}}
        \label{fig::aortic_valve::streamlines::pressure::3}
    \end{subfigure}
   \hspace{-2.7em}
    \begin{subfigure}[b]{0.48\textwidth}
        \centering
        \includegraphics[height=0.29\textwidth]{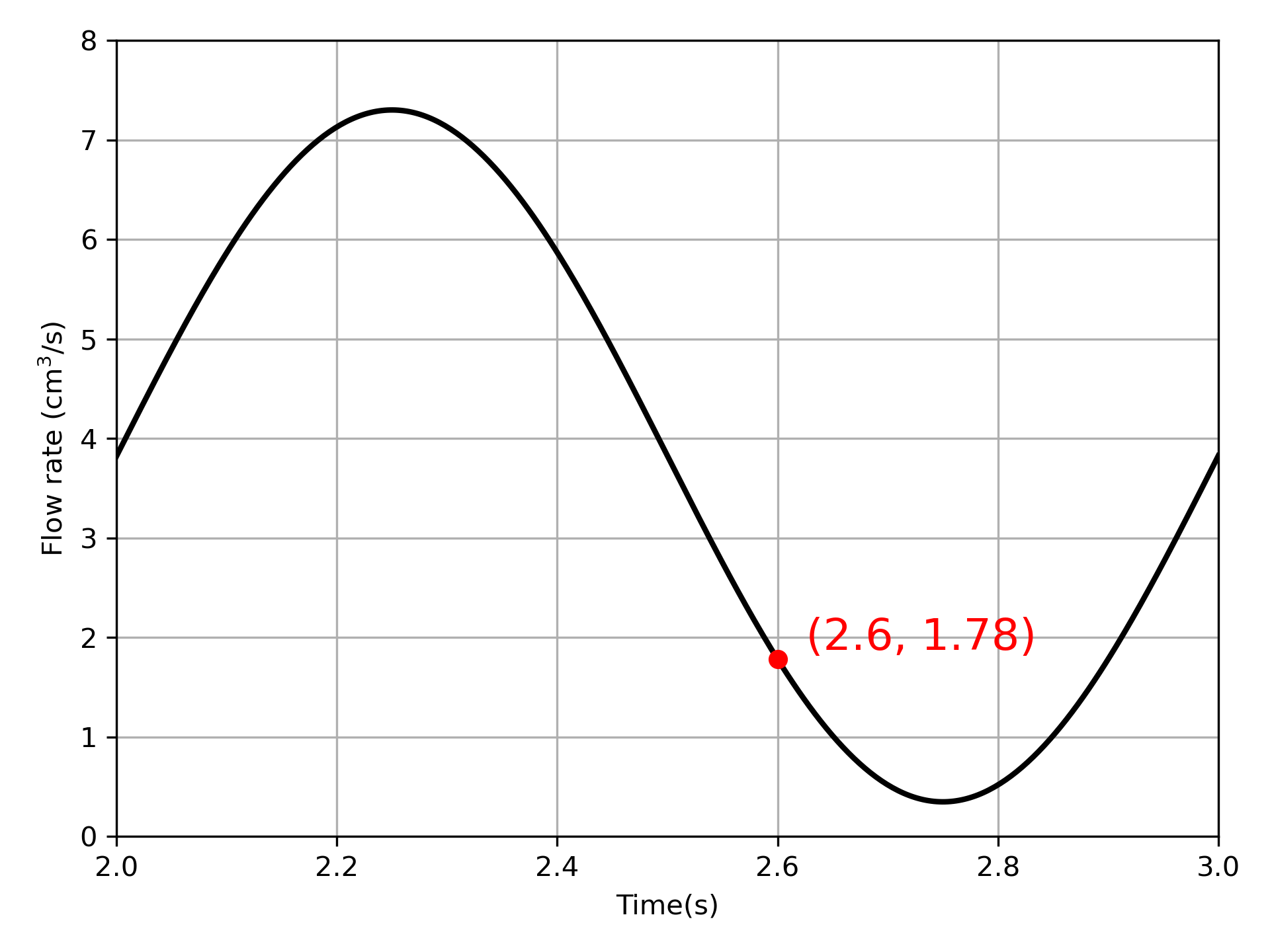}
        \caption{2.6~\text{s}}
        \label{fig::aortic_valve::streamlines::velocity::4}
    \end{subfigure}
    \hspace{-2.7em}
    \begin{subfigure}[b]{0.48\textwidth}
        \centering
        \includegraphics[height=0.29\textwidth]{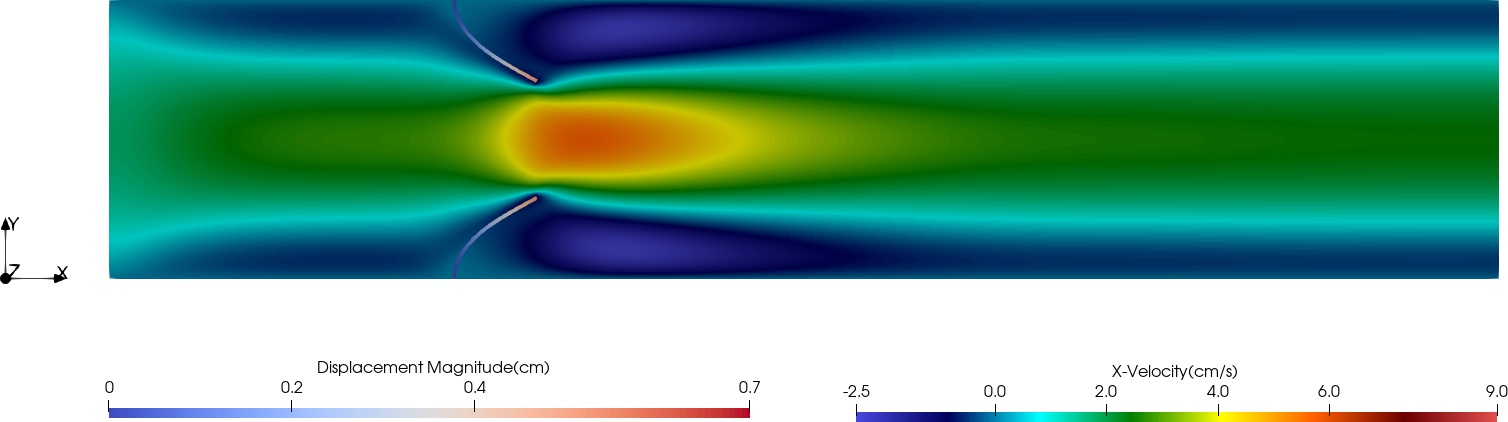}
        \caption{2.6~\text{s}}
        \label{fig::aortic_valve::streamlines::pressure::4}
    \end{subfigure}
    \begin{subfigure}[b]{0.48\textwidth}
        \centering
        \includegraphics[height=0.29\textwidth]{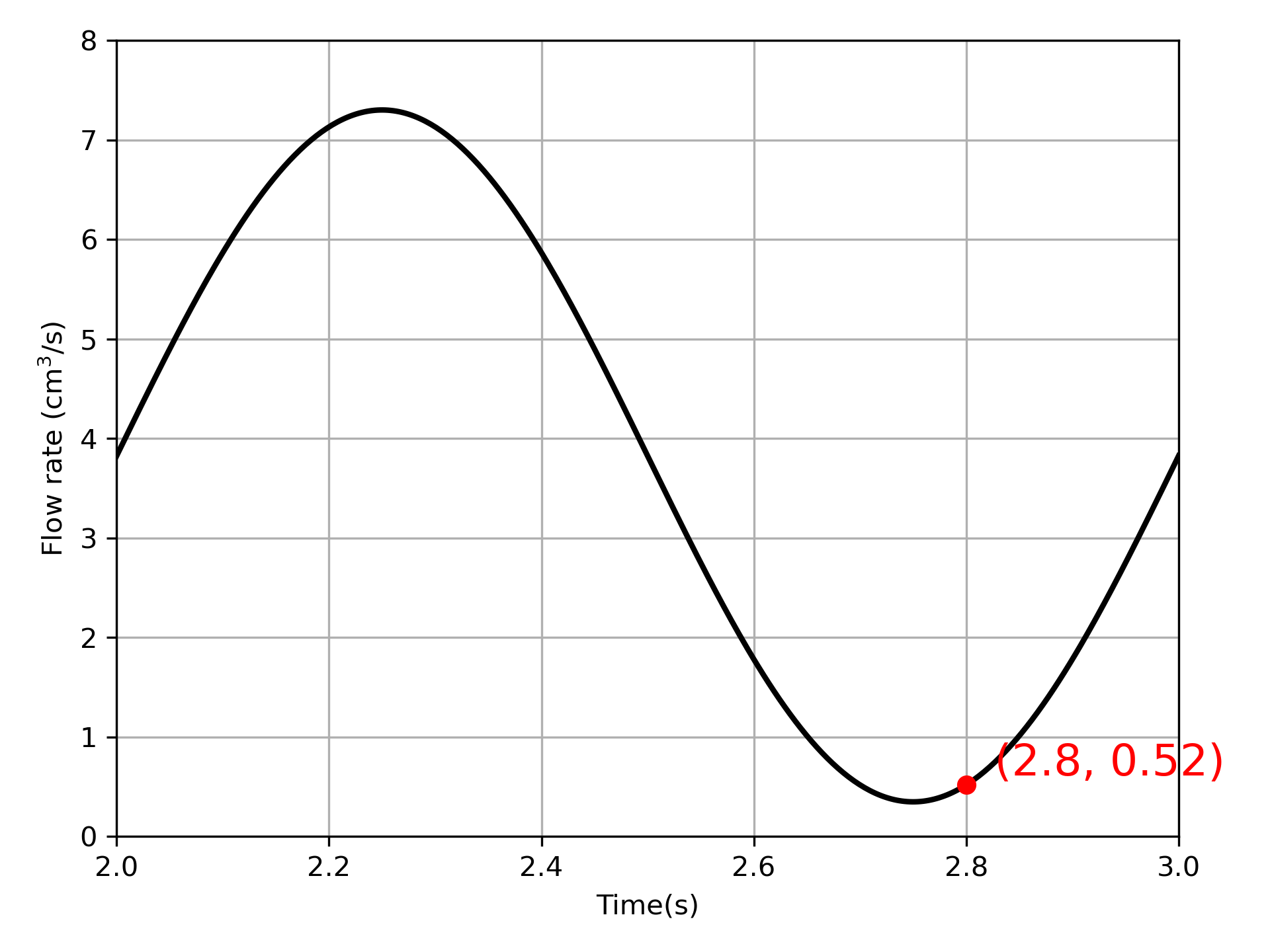}
        \caption{2.8~\text{s}}
        \label{fig::aortic_valve::streamlines::velocity::5}
    \end{subfigure}
    \hspace{-2.7em}
    \begin{subfigure}[b]{0.48\textwidth}
        \centering
        \includegraphics[height=0.29\textwidth]{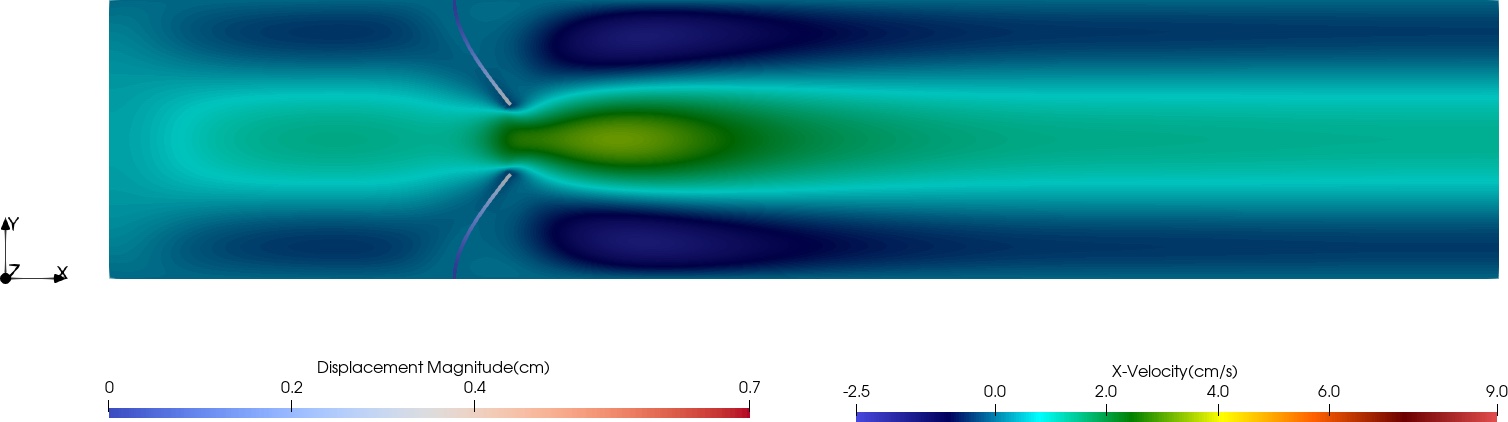}
        \caption{2.8~\text{s}}
        \label{fig::aortic_valve::streamlines::pressure::5}
    \end{subfigure}
    \caption{Fluid $x$-velocity and leaflet displacement magnitude evaluated at uniformly spaced 
        time points during the final simulated cardiac cycle. The left column presents 
        the corresponding flow rate, with red dots denoting the specific sampling instances.}
    \label{fig::Idealized_heart_valve::2D::streamlines}
\end{figure}

\begin{figure}[!ht]
    \centering
    \begin{minipage}{0.45\textwidth}
        \centering
        \includegraphics[width=1.0\textwidth]{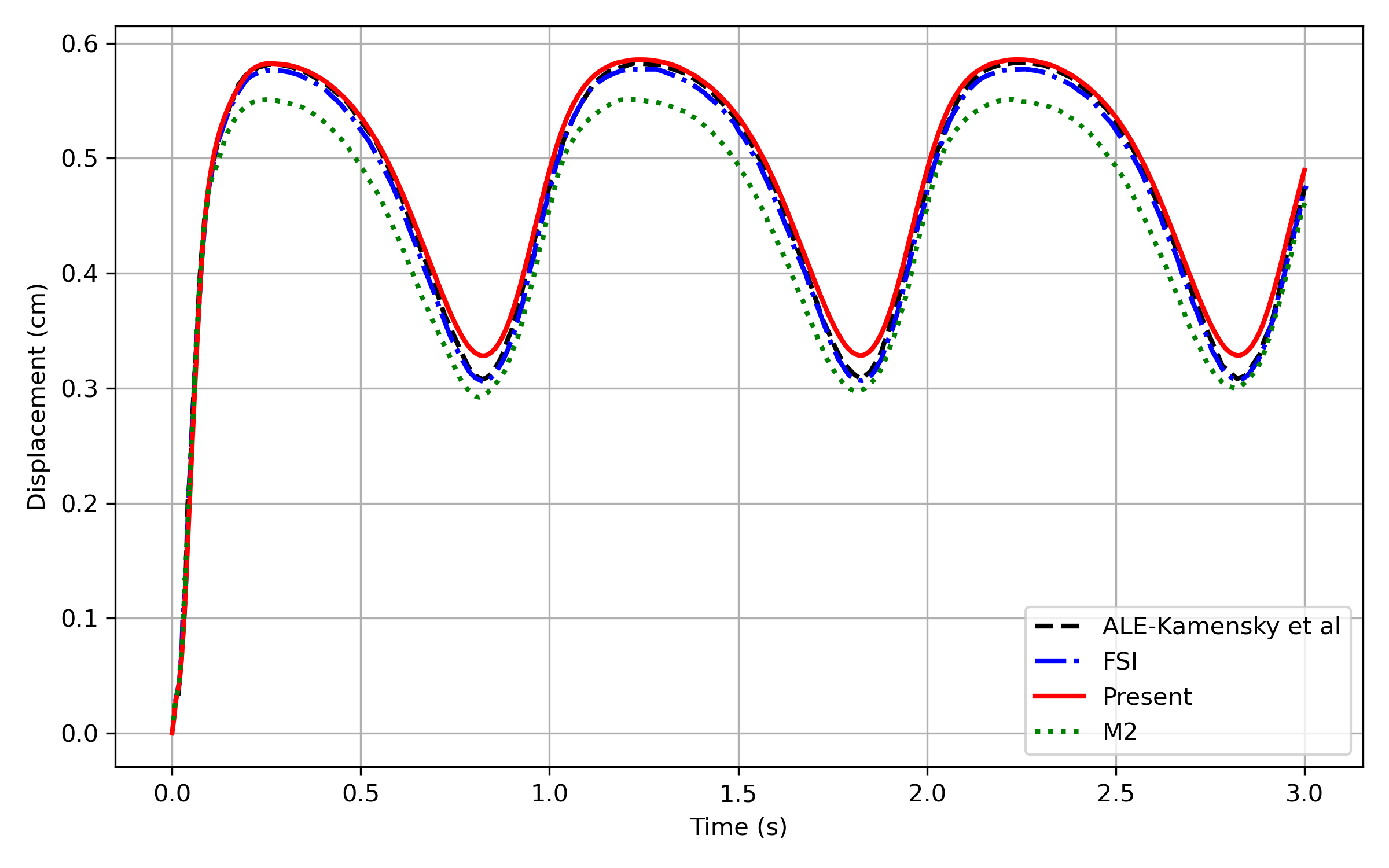}
        \subcaption{x}
        \label{fig::Idealized_heart_valve::2D::x}
    \end{minipage}
    \begin{minipage}{0.45\textwidth}
        \centering
        \includegraphics[width=1.0\textwidth]{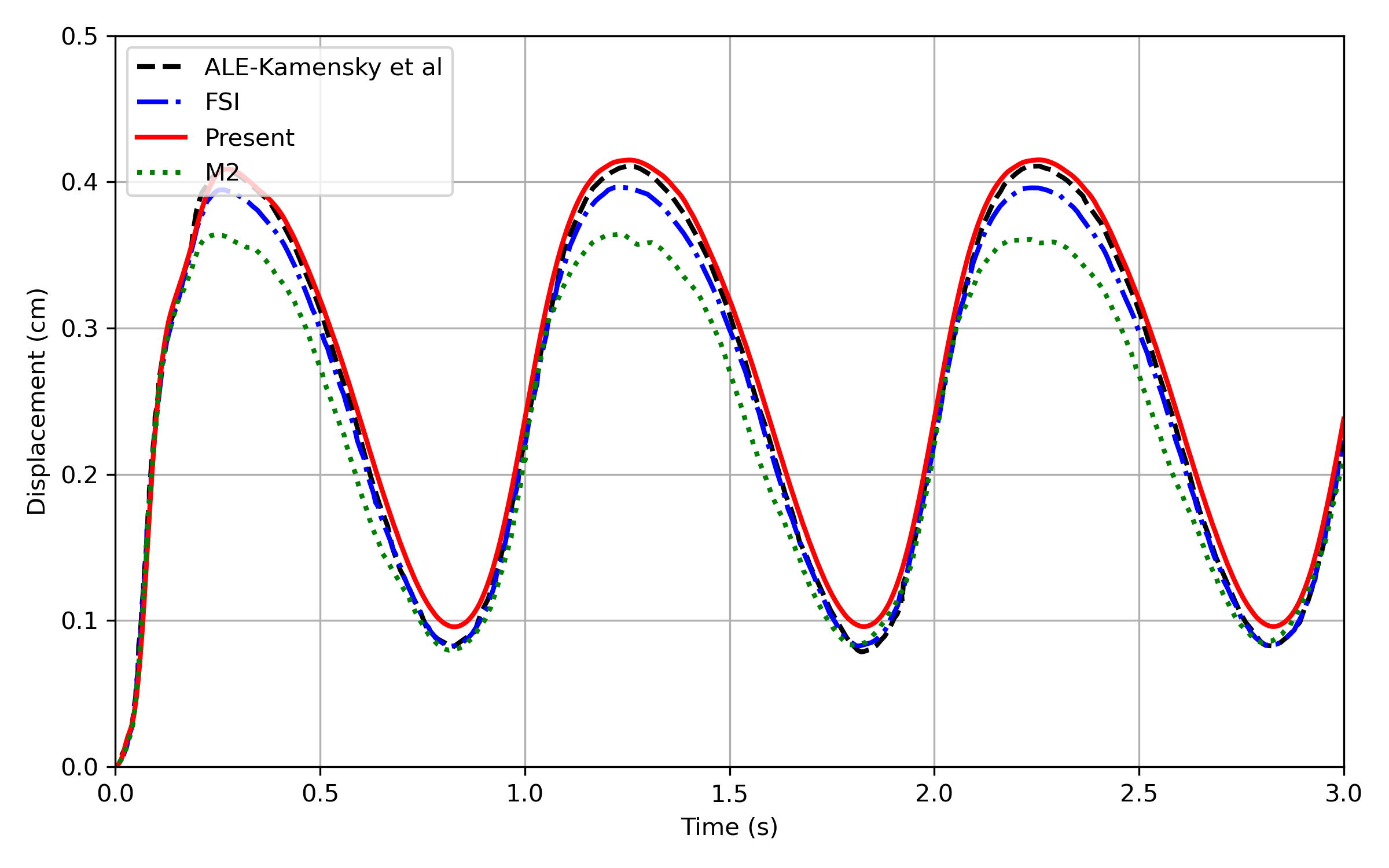}
        \subcaption{y}
        \label{fig::Idealized_heart_valve::2D::y}
    \end{minipage}
    \caption{$x$- and $y$-displacements of the tip of the top leaflet for the idealized heart valve model with isotropic properties.}
    \label{fig::Idealized_heart_valve::2D::displacement}
\end{figure}

In this case, the structural domain is discretized into $4\times40$ quadrilateral elements, 
while the fluid domain is discretized using a structured quadrilateral mesh of size $256\times64$.
Figure \ref{fig::Idealized_heart_valve::2D::streamlines} shows the velocity component 
in the x-direction and the displacement magnitude of the valve leaflets at different times 
during the cardiac cycle from 2 s to 3 s. 
The results exhibit excellent agreement with the simulations reported in \cite{black2025immersed}. 
As illustrated in the figure, when the flow rate approaches its peak, 
a high-velocity jet forms on the upstream surface of the valve, 
while a flow stagnation region develops immediately downstream of the leaflets.
Figure \ref{fig::Idealized_heart_valve::2D::displacement} presents a comparison of the time-dependent displacements 
at the tip of the upper valve leaflet in the 
$x$ and $y$ directions. The results from this study are compared with those obtained using the ALE-FSI method 
\cite{kamensky2015immersogeometric} and the distributed Lagrange multiplier (DLM) type FSI method reported 
in \cite{black2025immersed}. It is worth noting that the ALE-FSI method employs the Saint-Venant–Kirchhoff constitutive model, 
whereas the DLM method adopts the same material model as used in the present study. As shown in the figure, the displacement 
curves produced by the proposed 
method exhibit excellent agreement with both reference solutions, thereby validating the accuracy of the current approach.

\subsubsection{FSI validation: leaflets with anisotropic constitutive Materials}
In reality, heart valves are biologically complex and compositionally heterogeneous 
tissues that exhibit pronounced anisotropic mechanical behavior. 
To more accurately capture their biomechanical characteristics, the second case models 
the valve leaflets as anisotropic hyperelastic materials, employing the anisotropic 
constitutive model proposed in \cite{black2025immersed}, whose specific formulation is given below,
\begin{align}\label{eq::heart::valve::anisotropic}
    \Psi=\frac{C_{0}}{2}\left(\bar{I}_{1}-3\right)+C_{1}\left(\text{exp}(\bar{I}_{4f}-1)-\bar{I}_{4f}\right)+\frac{\kappa_{s}}{2}\left(\frac{1}{2}\left(J^{2}-1\right)-\text{ln}\left(J\right)\right)
\end{align}
where $C_{0}$ and $C_{1}$ are the material parameters. The first term in \eqref{eq::heart::valve::anisotropic} captures the isotropic response of the matrix, 
whereas the second term accounts for a nonlinear stiffening effect along a preferred direction, 
resulting in an overall transversely isotropic mechanical behavior.
The parameter settings used for The anisotropic material model are listed in Table \ref{tab::heart::valve::params::anisotropic}.
\begin{table}[!h]
    \centering
    \begin{tabular}{l|c|c}
        \toprule
        \textbf{Symbol} & \textbf{Value} & \textbf{Unit}  \\
        \midrule
        $T$     & 3.0   & $\text{s}$ \\
        $\rho$  & 100   & $\frac{\text{g}}{\text{cm}^3}$ \\
        $\mu$   & 10.0  & $\text{g}/(\text{cm}\cdot\text{s})$\\
        $C_{0}$  & $2.0\times10^{7}$  & $\text{Pa}$\\
        $C_{1}$  & $5\times C_{0}$  & $\text{Pa}$\\
        $\kappa_{s}$  & $1.0\times10^{7}$ & $\text{Pa}$\\
        $\Delta t$ & $1\times 10^{-4}$ & $\text{s}$\\
        \bottomrule
    \end{tabular}
    \caption{Parameters for the idealized heart valve model with anisotropic properties.}
    \label{tab::heart::valve::params::anisotropic}
\end{table}

\begin{figure}[!ht]
    \centering
    \includegraphics[width=0.4\textwidth]{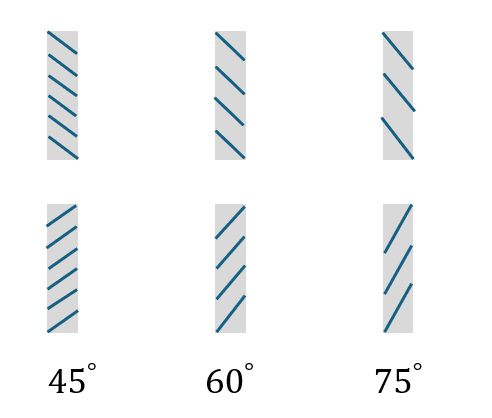}
    \caption{Top and bottom leaflets with different fiber orientations. The blue lines indicate the fiber directions. Note that the valve geometry is not drawn to scale.}
    \label{fig::Idealized_heart_valve::2D::fiber_orientation}
\end{figure}

\begin{figure}[!ht]
    \centering
    \begin{minipage}{0.45\textwidth}
        \centering
        \includegraphics[width=1.0\textwidth]{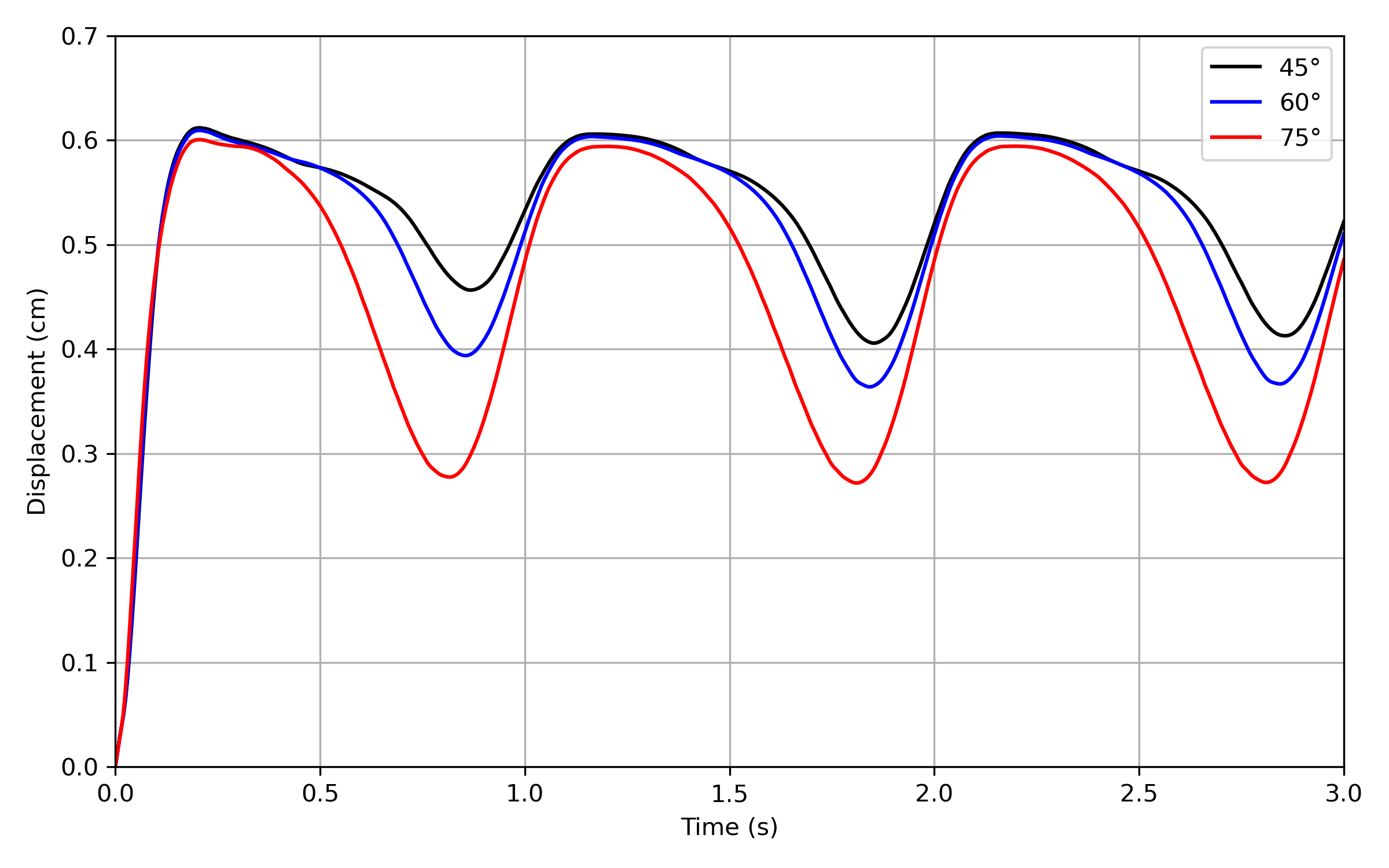}
        \subcaption{x}
        \label{fig::Idealized_heart_valve::2D::x}
    \end{minipage}
    \begin{minipage}{0.45\textwidth}
        \centering
        \includegraphics[width=1.0\textwidth]{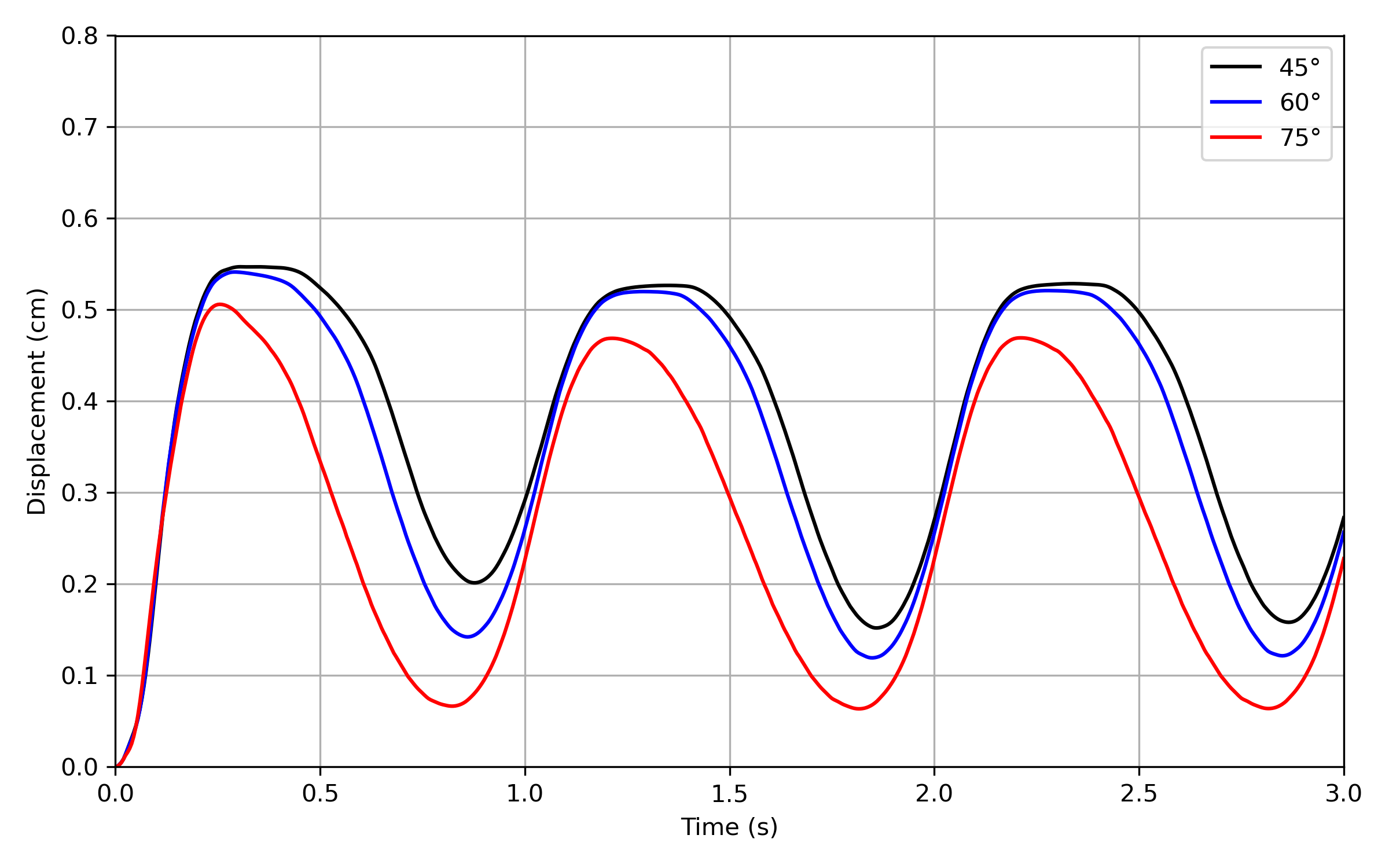}
        \subcaption{y}
        \label{fig::Idealized_heart_valve::2D::y}
    \end{minipage}
    \caption{$x$- and $y$-displacements of the tip of the top leaflet for the idealized heart valve model with anisotropic properties.}
    \label{fig::anisotropic::Idealized_heart_valve::2D}
\end{figure}

\begin{figure}[!ht]
    \centering
    \begin{subfigure}[c]{0.23\textwidth}
        \centering
        \includegraphics[width=\textwidth, height=3cm, keepaspectratio]{figure_software/2D_heart_valve_x2.00.png}
        \vspace{0.5cm}
        \label{fig:sub1}
    \end{subfigure}
    \begin{subfigure}[c]{0.23\textwidth}
        \centering
        \includegraphics[width=\textwidth, height=4.5cm, keepaspectratio]{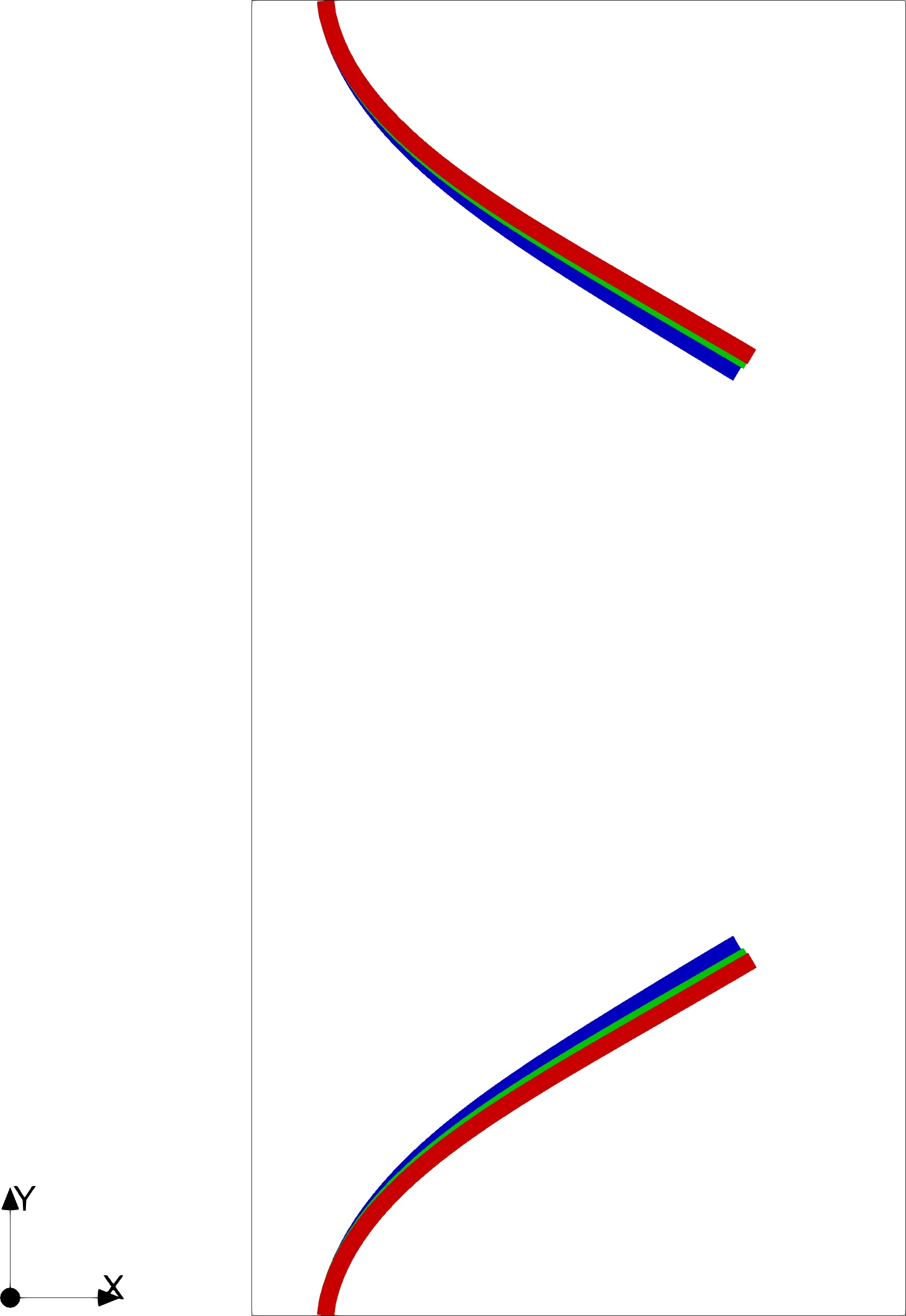}
        \vspace{0.5cm}
        \label{fig:sub2}
    \end{subfigure}
    \begin{subfigure}[c]{0.23\textwidth}
        \centering
        \includegraphics[width=\textwidth, height=3cm, keepaspectratio]{figure_software/2D_heart_valve_x2.20.png}
        \vspace{0.5cm}
        \label{fig:sub3}
    \end{subfigure}
    \begin{subfigure}[c]{0.23\textwidth}
        \centering
        \includegraphics[width=\textwidth, height=4.5cm, keepaspectratio]{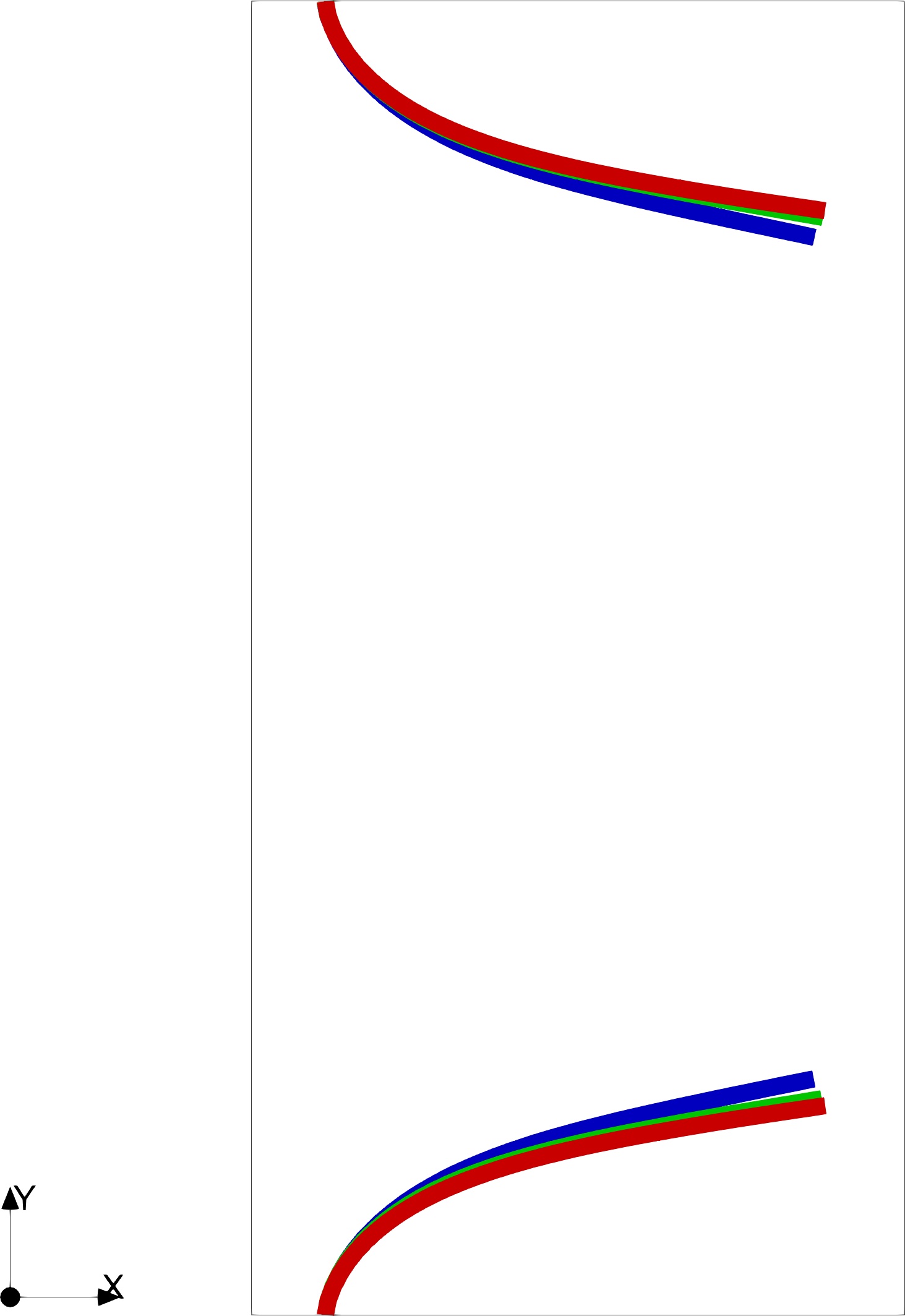}
        \vspace{0.5cm}
        \label{fig:sub4}
    \end{subfigure}
    \begin{subfigure}[c]{0.23\textwidth}
        \centering
        \includegraphics[width=\textwidth, height=3cm, keepaspectratio]{figure_software/2D_heart_valve_x2.40.png}
        \vspace{0.5cm}
        \label{fig:sub5}
    \end{subfigure}
    \begin{subfigure}[c]{0.23\textwidth}
        \centering
        \includegraphics[width=\textwidth, height=4.5cm, keepaspectratio]{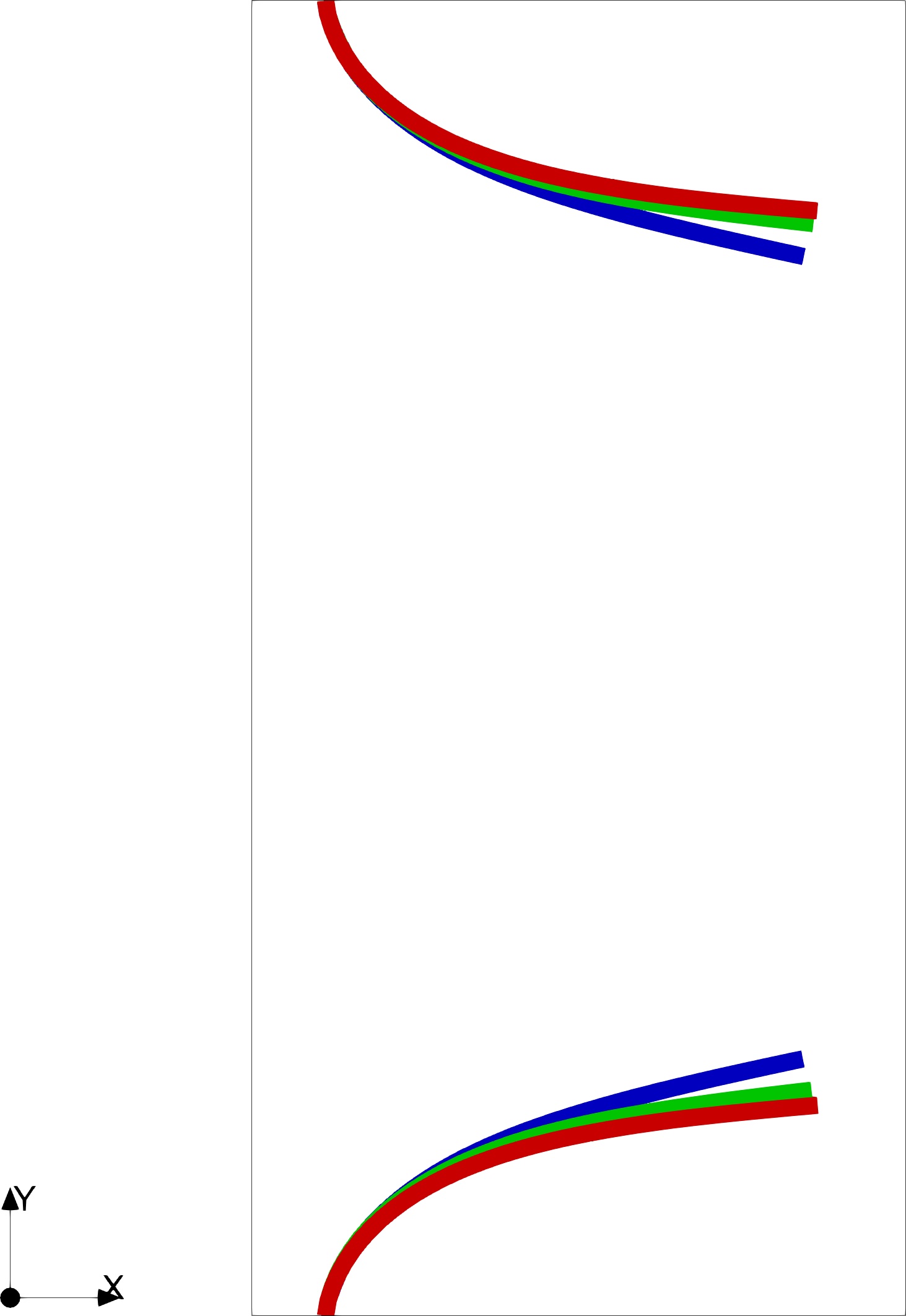}
        \vspace{0.5cm}
        \label{fig:sub6}
    \end{subfigure}
    \begin{subfigure}[c]{0.23\textwidth}
        \centering
        \includegraphics[width=\textwidth, height=3cm, keepaspectratio]{figure_software/2D_heart_valve_x2.60.png}
        \vspace{0.5cm}
        \label{fig:sub7}
    \end{subfigure}
    \begin{subfigure}[c]{0.23\textwidth}
        \centering
        \includegraphics[width=\textwidth, height=4.5cm, keepaspectratio]{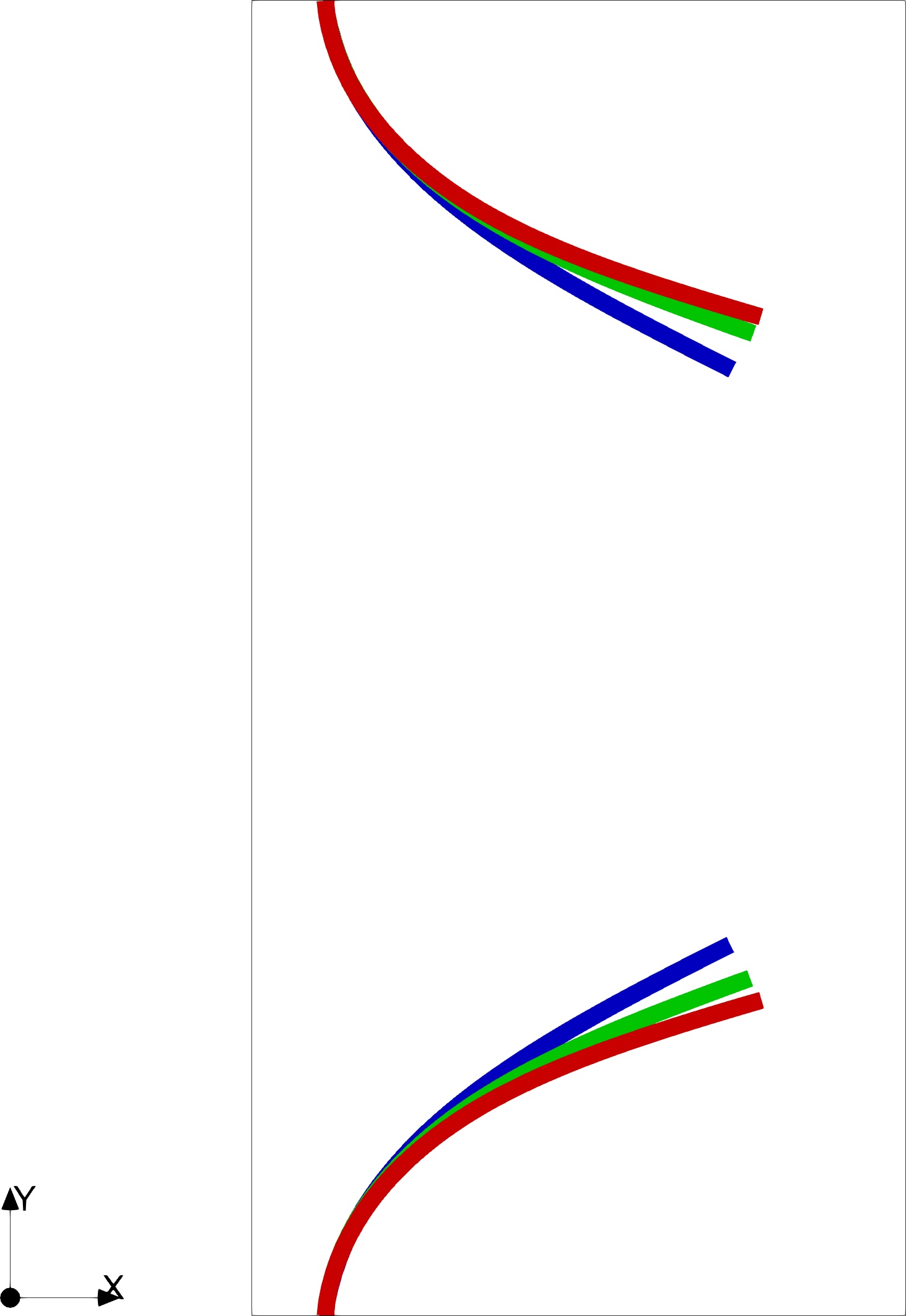}
        \vspace{0.5cm}
        \label{fig:sub8}
    \end{subfigure}
    \begin{subfigure}[c]{0.23\textwidth}
        \centering
        \includegraphics[width=\textwidth, height=3cm, keepaspectratio]{figure_software/2D_heart_valve_x2.80.png}
        \vspace{0.5cm}
        \label{fig:sub9}
    \end{subfigure}
    \begin{subfigure}[c]{0.23\textwidth}
        \centering
        \includegraphics[width=\textwidth, height=4.5cm, keepaspectratio]{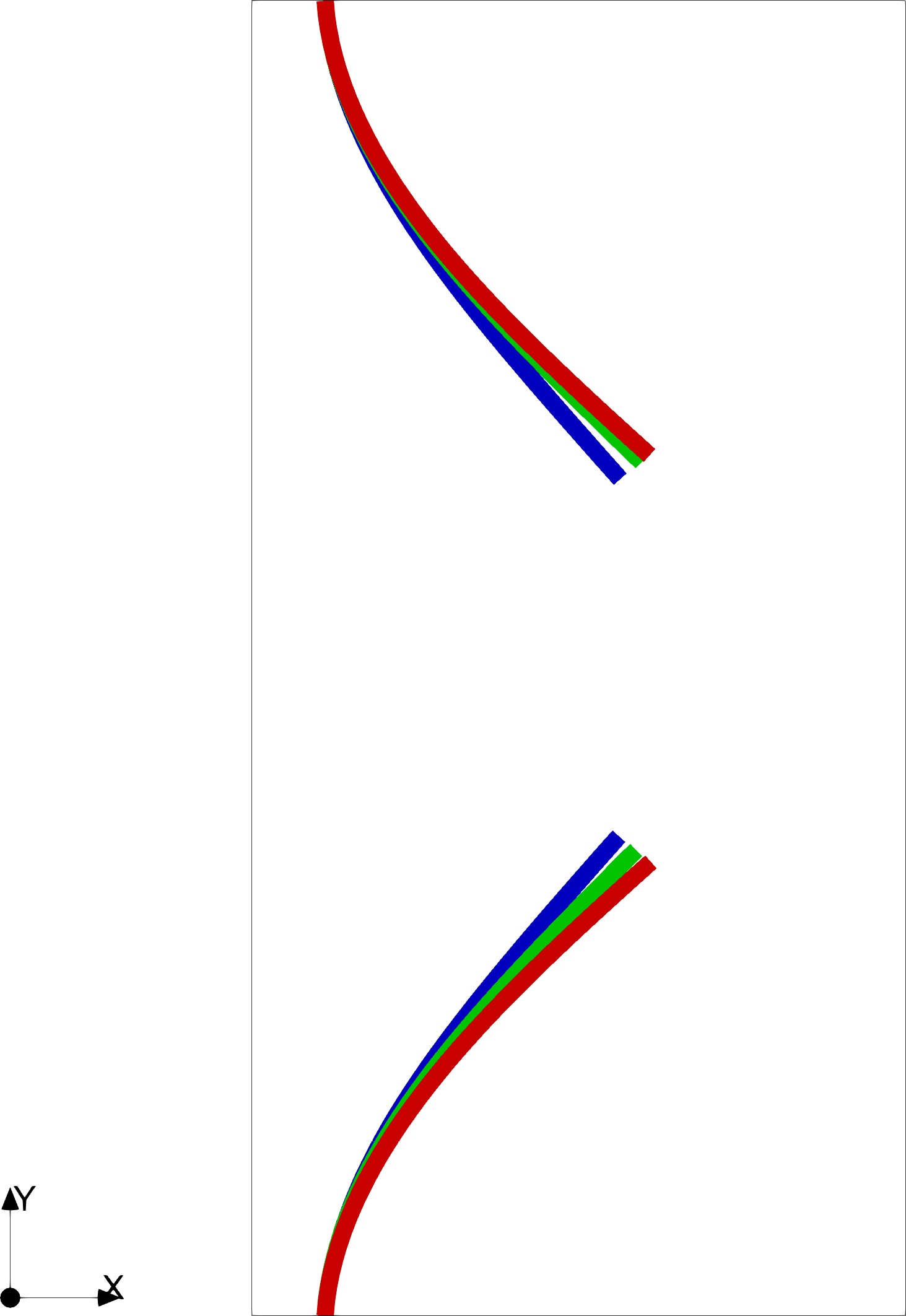}
        \vspace{0.5cm}
        \label{fig:sub10}
    \end{subfigure}
    \begin{subfigure}[c]{0.46\textwidth}
        \centering
        \includegraphics[height=3.0cm]{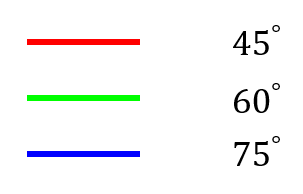}
        \vspace{0.5cm}
        \label{fig:sub11}
    \end{subfigure}
    \caption{Deformation of the leaflets during the final simulated heartbeat. Each color represents a different fiber orientation.}
    \label{fig::anisotropic::Idealized_heart_valve::2D::deformation}
\end{figure}

In this case, the structural and fluid domains adopt the same mesh configuration as in the first case.
Three different fiber orientation angles ($45^\circ$, $60^\circ$, and $75^\circ$), 
as illustrated in Figure \ref{fig::Idealized_heart_valve::2D::fiber_orientation}, 
are considered to assess the capability of the AFSI software in modeling anisotropic constitutive materials.
The simulation results indicate that the chosen material properties and constitutive model produce 
a more compliant isotropic background matrix in the leaflets, leading to more complex deformation 
at certain fiber orientations.
Figure~\ref{fig::anisotropic::Idealized_heart_valve::2D} shows the displacements of the upper leaflet tip 
in the $x$ and $y$ directions during the final cardiac cycle.
Figure~\ref{fig::anisotropic::Idealized_heart_valve::2D::deformation} illustrates the motion of the valve 
leaflets with different fiber orientations over the same period.
As shown, the displacement and overall dynamic behavior of the leaflet with a $60^\circ$ fiber orientation 
lie between those observed at $45^\circ$ and $75^\circ$.
Among the three cases, the $45^\circ$ orientation yields the largest leaflet tip displacement, 
while the $75^\circ$ orientation results in the smallest.
These results underscore the significant role of fiber orientation in influencing the mechanical 
response of the valve leaflets, which is crucial for evaluating valve performance under different 
physiological conditions.
\subsection{Idealized left ventricle}

\begin{figure}[!ht]
    \centering
    \includegraphics[width=0.4\textwidth]{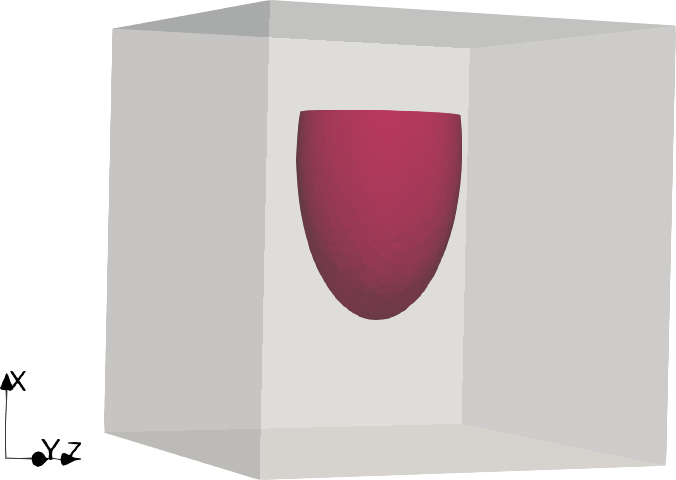}
    \caption{Setup of the idealized left ventricle in the IB framework: the gray region represents the fixed computational domain $\Omega$, while the 
            red region corresponds to the deformable idealized left ventricle.}
    \label{fig::Idealized_left_ventricle::geometry}
\end{figure}

To demonstrate the numerical simulation capabilities of the AFSI software in handling complex geometries and structural models, 
a three-dimensional idealized left ventricle is employed as a test case. This model serves as a simplified representation 
of the human left ventricle and is widely used as a benchmark for verifying the accuracy of cardiac mechanics solvers 
\cite{land2015verification,griffith2017hybrid}. 
Figure~\ref{fig::Idealized_left_ventricle::geometry} depicts the initial geometry of the idealized 
left ventricle, along with the associated fixed computational domain. The undeformed geometry is defined 
through the following parameterization of a truncated ellipsoid:
\begin{align*}
    \mathbf{X}(u,v) = 
    \begin{pmatrix}
    x \\
    y \\
    z
    \end{pmatrix}
    =
    \begin{pmatrix}
    r_s \sin u \cos v \\
    r_s \sin u \sin v \\
    r_l \cos u
    \end{pmatrix}.
\end{align*}

The boundary parameters are defined as follows:
\begin{itemize}
 \item Endocardial surface: $r_{s}=0.7~\text{cm}$, $r_{l}=1.7~\text{cm}$, $u\in\left[-\pi,-\arccos\frac{5}{17}\right]$, and $v\in[-\pi,\pi]$;
 \item Epicardial surface: $r_{s}=1~\text{cm}$, $r_{l}=2.0~\text{cm}$, $u\in[-\pi,-arccos\frac{5}{20}]$, and $v\in[-\pi,\pi]$;
 \item Base plane: $z=0.5~\text{cm}$, which is implicitly defined by the given range of $u$.
\end{itemize}

The passive mechanical behavior of the left ventricle is modeled using a transversely isotropic 
hyperelastic constitutive law proposed by Guccione et al.~\cite{guccione1995finite}:
\begin{align}
    \Psi = \frac{C}{2}\left(\text{exp}(Q)-1\right)
\end{align}
where
\begin{align*}
 Q=b_{f}E_{11}^{2}+b_{t}(E_{22}^{2}+E_{33}^{2}+E_{23}^{2}+E_{32}^{2})+b_{fs}(E_{12}^{2}+E_{21}^{2}+E_{13}^{2}+E_{31}^{2}),
\end{align*}
and $\mathbb{E}=\frac{1}{2}(\mathbb{F}^{T}\mathbb{F}-\mathbb{I})=(E_{ij})$ is the Green-Lagrange 
strain tensor. Here, $C$, $b_{f}$, $b_{t}$, and $b_{fs}$ are material parameters.
The geometry of the idealized left ventricle is discretized using 17,879 tetrahedral elements, 
and the fluid computational domain $\Omega$ is resolved on a $64 \times 64 \times 64$ Cartesian mesh. 
The parameters employed in the simulation are summarized in Table~\ref{tab::left::ventricle::params}.
A uniform pressure load of $10~\text{kPa}$ is applied to the endocardial surface to simulate 
physiological loading conditions.

\begin{table}[!h]
    \centering
    \begin{tabular}{l|c|c}
        \toprule
        \textbf{Symbol} & \textbf{Value} & \textbf{Unit}  \\
        \midrule
        $T$        & 3.0  & $\text{s}$ \\
        $\rho$     & 1    & $\frac{\text{g}}{\text{cm}^3}$ \\
        $\mu$      & 0.04 & $\text{g}/(\text{cm}\cdot\text{s})$\\
        $C$        & $10$ & $\text{kPa}$\\
        $b_{f}$    & $1$  & -\\
        $b_{t}$    & $1$  & -\\
        $b_{fs}$   & $1$  & -\\
        $\Delta t$ & $1\times 10^{-4}$ & $\text{s}$\\
        \bottomrule
    \end{tabular}
    \caption{Parameters for the idealized left ventricle model.}
    \label{tab::left::ventricle::params}
\end{table}

\section{Conclusions} \label{sec:conclusions}
This implementation enables seamless integration of the IB method within the FEniCS framework for FSI simulations, offering several notable advantages.
First, the nodal coupling scheme naturally accommodates arbitrary high-order finite element spaces, 
ensuring both flexibility in spatial discretization and high-order accuracy. 
Second, the methodology completely avoids the need for remeshing when handling large deformations and displacements, 
thereby substantially reducing preprocessing complexity and computational overhead. 
Third, the framework permits straightforward selection among various time discretization schemes 
and solution algorithms through FEniCS's native time-stepping capabilities. 
Finally, the implementation supports automatic derivation and implementation of 
finite element formulations directly from strain energy functions, 
which particularly benefits modeling of solids with complex constitutive behavior and significantly streamlines the development of new material models.

\section*{Author contributions}
\textbf{Pengfei Ma:} Methodology (lead); Software(lead); Writing - original draft (lead).
\textbf{Li Cai:} Supervision (lead); Writing - review \& editing (equal).
\textbf{Xuan Wang:} Visualization (lead); Writing - original draft (supporting); Writing - review \& editing (equal). 
\textbf{Hao Gao:}  Writing - review \& editing (equal).
\section*{Declaration of competing interest}
The authors declare that they have no known competing financial interests or personal relationships that could have appeared to influence the work reported in this paper.

\section*{Data availability}
Data will be made available on request.

\section*{Acknowledgements}
Pengfei Ma, Li Cai and Xuan Wang were supported by the National Natural Science Foundation of China (Grant No. 12271440).  
Hao Gao was supported by the British Heart Foundation (PG/22/10930) and the Engineering and Physical Sciences Research Council (EPSRC) of the United Kingdom (EP/S030875/1). 

\newpage

\bibliographystyle{unsrt}
\bibliography{references}  

\begin{thebibliography}{10}

\bibitem{lee_fluidstructure_2020}
Jae~H. Lee, Alex~D. Rygg, Ebrahim~M. Kolahdouz, Simone Rossi, Stephen~M. Retta, Nandini Duraiswamy, Lawrence~N. Scotten, Brent~A. Craven, and Boyce~E. Griffith.
\newblock Fluid–{Structure} {Interaction} {Models} of {Bioprosthetic} {Heart} {Valve} {Dynamics} in an {Experimental} {Pulse} {Duplicator}.
\newblock {\em Annals of Biomedical Engineering}, 48(5):1475--1490, May 2020.

\bibitem{bavo_fluid-structure_2016}
Alessandra~M. Bavo, Giorgia Rocatello, Francesco Iannaccone, Joris Degroote, Jan Vierendeels, and Patrick Segers.
\newblock Fluid-{Structure} {Interaction} {Simulation} of {Prosthetic} {Aortic} {Valves}: {Comparison} between {Immersed} {Boundary} and {Arbitrary} {Lagrangian}-{Eulerian} {Techniques} for the {Mesh} {Representation}.
\newblock {\em PLOS ONE}, 11(4):e0154517, April 2016.

\bibitem{lin_fluidstructure_2019}
Zhaowu Lin, Andrew Hess, Zhaosheng Yu, Shengqiang Cai, and Tong Gao.
\newblock A fluid–structure interaction study of soft robotic swimmer using a fictitious domain/active-strain method.
\newblock {\em Journal of Computational Physics}, 376:1138--1155, January 2019.

\bibitem{kuchumov_fluidstructure_2023}
Alex~G. Kuchumov, Anastasiya Makashova, Sergey Vladimirov, Vsevolod Borodin, and Anna Dokuchaeva.
\newblock Fluid–{Structure} {Interaction} {Aortic} {Valve} {Surgery} {Simulation}: {A} {Review}.
\newblock {\em Fluids}, 8(11):295, November 2023.

\bibitem{borazjani_review_2015}
Iman Borazjani.
\newblock A {Review} of {Fluid}-{Structure} {Interaction} {Simulations} of {Prosthetic} {Heart} {Valves}.
\newblock {\em Journal of Long-Term Effects of Medical Implants}, 25(1-2):75--93, 2015.

\bibitem{wang_development_2022}
Rui Wang, Shuo Wang, Yu~Wang, Long Cheng, and Min Tan.
\newblock Development and {Motion} {Control} of {Biomimetic} {Underwater} {Robots}: {A} {Survey}.
\newblock {\em IEEE Transactions on Systems, Man, and Cybernetics: Systems}, 52(2):833--844, February 2022.

\bibitem{li_underwater_2023}
Gongbo Li, Guijie Liu, Dingxin Leng, Xin Fang, Guanghao Li, and Wenqian Wang.
\newblock Underwater {Undulating} {Propulsion} {Biomimetic} {Robots}: {A} {Review}.
\newblock {\em Biomimetics}, 8(3):318, July 2023.

\bibitem{qu_recent_2024}
Juntian Qu, Yining Xu, Zhenkun Li, Zhenping Yu, Baijin Mao, Yunfei Wang, Ziqiang Wang, Qigao Fan, Xiang Qian, Min Zhang, Minyi Xu, Bin Liang, Houde Liu, Xueqian Wang, Xiaohao Wang, and Tiefeng Li.
\newblock Recent {Advances} on {Underwater} {Soft} {Robots}.
\newblock {\em Advanced Intelligent Systems}, 6(2):2300299, February 2024.

\bibitem{pramanik_computational_2024}
R.~Pramanik, R.~W. C.~P. Verstappen, and P.~R. Onck.
\newblock Computational fluid–structure interaction in biology and soft robots: {A} review.
\newblock {\em Physics of Fluids}, 36(10):101302, October 2024.

\bibitem{abbas_state_art_2022}
Syed~Samar Abbas, Mohammad~Shakir Nasif, and Rafat Al-Waked.
\newblock State-of-the-art numerical fluid–structure interaction methods for aortic and mitral heart valves simulations: {A} review.
\newblock {\em SIMULATION}, 98(1):3--34, January 2022.

\bibitem{bergersen_turtlefsi_2020}
Aslak Bergersen, Andreas Slyngstad, Sebastian Gjertsen, Alban Souche, and Kristian Valen-Sendstad.
\newblock {turtleFSI}: {A} {Robust} and {Monolithic} {FEniCS}-based {Fluid}-{Structure} {Interaction} {Solver}.
\newblock {\em Journal of Open Source Software}, 5(50):2089, June 2020.

\bibitem{bavo_fluid_structure_2016}
Alessandra~M. Bavo, Giorgia Rocatello, Francesco Iannaccone, Joris Degroote, Jan Vierendeels, and Patrick Segers.
\newblock Fluid-{Structure} {Interaction} {Simulation} of {Prosthetic} {Aortic} {Valves}: {Comparison} between {Immersed} {Boundary} and {Arbitrary} {Lagrangian}-{Eulerian} {Techniques} for the {Mesh} {Representation}.
\newblock {\em PLOS ONE}, 11(4):e0154517, April 2016.

\bibitem{boffi_hyper_elastic_2008}
Daniele Boffi, Lucia Gastaldi, Luca Heltai, and Charles~S. Peskin.
\newblock On the hyper-elastic formulation of the immersed boundary method.
\newblock {\em Computer Methods in Applied Mechanics and Engineering}, 197(25-28):2210--2231, April 2008.

\bibitem{feng_whole_heart_2024}
Liuyang Feng, Hao Gao, and Xiaoyu Luo.
\newblock Whole-heart modelling with valves in a fluid–structure interaction framework.
\newblock {\em Computer Methods in Applied Mechanics and Engineering}, 420:116724, February 2024.

\bibitem{griffith_hybrid_2017}
Boyce~E Griffith and Xiaoyu Luo.
\newblock Hybrid finite difference/finite element immersed boundary method.
\newblock {\em International Journal for Numerical Methods in Biomedical Engineering}, 33(12):e2888, December 2017.

\bibitem{wells_nodal_2023}
David~R. Wells, Ben Vadala-Roth, Jae~H. Lee, and Boyce~E. Griffith.
\newblock A nodal immersed finite element-finite difference method.
\newblock {\em Journal of Computational Physics}, 477:111890, March 2023.

\bibitem{boffi2008hyper}
Daniele Boffi, Lucia Gastaldi, Luca Heltai, and Charles~S Peskin.
\newblock On the hyper-elastic formulation of the immersed boundary method.
\newblock {\em Computer Methods in Applied Mechanics and Engineering}, 197(25-28):2210--2231, 2008.

\bibitem{vadala2020stabilization}
Ben Vadala-Roth, Shashank Acharya, Neelesh~A Patankar, Simone Rossi, and Boyce~E Griffith.
\newblock Stabilization approaches for the hyperelastic immersed boundary method for problems of large-deformation incompressible elasticity.
\newblock {\em Computer Methods in Applied Mechanics and Engineering}, 365:112978, 2020.

\bibitem{gerhard2000nonlinear}
A~Holzapfel Gerhard.
\newblock Nonlinear solid mechanics: A continuum approach for engineering, 2000.

\bibitem{Langtangen2012}
Anders Logg, Kent-Andre Mardal, and Garth Wells.
\newblock {\em A FEniCS tutorial}, pages 1--73.
\newblock Springer Berlin Heidelberg, Berlin, Heidelberg, 2012.

\bibitem{nanobind}
Wenzel Jakob.
\newblock nanobind: tiny and efficient c++/python bindings, 2022.
\newblock https://github.com/wjakob/nanobind.

\bibitem{wang_massively_2020}
Xinlei Wang, Yuxing Qiu, Stuart~R. Slattery, Yu~Fang, Minchen Li, Song-Chun Zhu, Yixin Zhu, Min Tang, Dinesh Manocha, and Chenfanfu Jiang.
\newblock A massively parallel and scalable multi-{GPU} material point method.
\newblock {\em ACM Transactions on Graphics}, 39(4), August 2020.

\bibitem{black2025immersed}
Ryan~T Black and George~Ilhwan Park.
\newblock An immersed fluid--structure interaction method targeted for heart valve applications.
\newblock {\em Computer Methods in Applied Mechanics and Engineering}, 435:117634, 2025.

\bibitem{kamensky2015immersogeometric}
David Kamensky, Ming-Chen Hsu, Dominik Schillinger, John~A Evans, Ankush Aggarwal, Yuri Bazilevs, Michael~S Sacks, and Thomas~JR Hughes.
\newblock An immersogeometric variational framework for fluid--structure interaction: Application to bioprosthetic heart valves.
\newblock {\em Computer Methods in Applied Mechanics and Engineering}, 284:1005--1053, 2015.

\bibitem{hesch2012continuum}
C~Hesch, AJ~Gil, A~Arranz Carreno, and J~Bonet.
\newblock On continuum immersed strategies for fluid--structure interaction.
\newblock {\em Computer Methods in Applied Mechanics and Engineering}, 247:51--64, 2012.

\bibitem{gil2010immersed}
Antonio~J Gil, A~Arranz Carre{\~n}o, Javier Bonet, and Oubay Hassan.
\newblock The immersed structural potential method for haemodynamic applications.
\newblock {\em Journal of Computational Physics}, 229(22):8613--8641, 2010.

\bibitem{land2015verification}
Sander Land, Viatcheslav Gurev, Sander Arens, Christoph~M Augustin, Lukas Baron, Robert Blake, Chris Bradley, Sebastian Castro, Andrew Crozier, Marco Favino, et~al.
\newblock Verification of cardiac mechanics software: benchmark problems and solutions for testing active and passive material behaviour.
\newblock {\em Proceedings of the Royal Society A: Mathematical, Physical and Engineering Sciences}, 471(2184):20150641, 2015.

\bibitem{griffith2017hybrid}
Boyce~E Griffith and Xiaoyu Luo.
\newblock Hybrid finite difference/finite element immersed boundary method.
\newblock {\em International Journal for Numerical Methods in Biomedical Engineering}, 33(12):e2888, 2017.

\bibitem{guccione1995finite}
Julius~M Guccione, Kevin~D Costa, and Andrew~D McCulloch.
\newblock Finite element stress analysis of left ventricular mechanics in the beating dog heart.
\newblock {\em Journal of Biomechanics}, 28(10):1167--1177, 1995.

\end{thebibliography}
\end{document}